%% file: sample-manuscript.tex
\documentclass[manuscript]{acmart} 

\usepackage{graphicx}
\usepackage{subcaption}
\usepackage{algorithm}
\usepackage{algorithmic}
\usepackage{enumitem}
\usepackage{color, colortbl}
\usepackage{tcolorbox} 

\usepackage{adjustbox}

\begin{document}

\title{Diffusion Language Model for Recommendation}

\author{Chengyi Liu}
\affiliation{%
  \institution{The Hong Kong Polytechnic University}
  \country{Hong Kong SAR}}
\email{chengyi.liu@connect.polyu.hk}

\author{Yongqi Zhou}
\affiliation{%
  \institution{Tencent Inc.}
  \country{Shenzhen, China}}
\email{kolinzhou@tencent.com}

\author{Junwei Pan}
\affiliation{%
  \institution{Tencent Inc.}
  \country{Shenzhen, China}}
\email{jonaspan@tencent.com}

\author{Zhixiang Feng}
\affiliation{%
  \institution{Tencent Inc.}
  \country{Shenzhen, China}}
\email{lionelfeng@tencent.com}

\author{Chengguo Yin}
\affiliation{%
  \institution{Tencent Inc.}
  \country{Shenzhen, China}}
\email{turingyin@tencent.com}

\author{Haijie Gu}
\affiliation{%
  \institution{Tencent Inc.}
  \country{Shenzhen, China}}
\email{jerrickgu@tencent.com}

\author{Jie Jiang}
\affiliation{%
  \institution{Tencent Inc.}
  \country{Shenzhen, China}}
\email{zeus@tencent.com}

\author{Yinghao Liu}
\affiliation{%
 \institution{The Hong Kong Polytechnic University}
 \country{Hong Kong SAR}}
\email{liuyinghao@shu.edu.cn}

\author{Yujuan Ding}
\affiliation{%
 \institution{The Hong Kong Polytechnic University}
 \country{Hong Kong SAR}}
\email{dingyujuan385@gmail.com}

\author{Qing Li*}
\affiliation{%
 \institution{The Hong Kong Polytechnic University}
 \country{Hong Kong SAR}}
\email{qing-prof.li@polyu.edu.hk}

\author{Wenqi  Fan}
\authornote{Corresponding author: Wenqi Fan and Qing Li, Department of Computing, The Hong Kong Polytechnic University}
\affiliation{%
 \institution{The Hong Kong Polytechnic University}
 \country{Hong Kong SAR}}
\email{wenqifan03@gmail.com}

\renewcommand{\shortauthors}{Liu et al.}

\begin{abstract}
Large language model (LLM)-empowered recommender systems have emerged as a promising paradigm for generative recommendation, leveraging their strong semantic reasoning and generative capacity to model complex, diverse user preferences.
However, most existing approaches rely on an autoregressive paradigm that is suboptimal for recommendation. The next-token objective emphasizes sequential order rather than the structural inter-item dependencies underlying user preferences. In addition, prefix-constrained generation restricts bidirectional context and commits to left-to-right decoding, causing early errors to accumulate without correction.
Inspired by the success of diffusion language models, we propose \textbf{DLMRec}, a discrete diffusion language model tailored for recommendation that offers a compelling alternative to autoregressive generation.
Specifically, DLMRec introduces three key components to bridge diffusion language modeling with recommendation. First, a collaborative-aware stochastic tokenizer encodes multi-hop collaborative signals into expressive discrete tokens compatible with diffusion modeling. Second, a curriculum-driven training strategy aligns the denoising process with preference recovery through progressive item- and token-level learning. Third, a stability-aware voting mechanism aggregates iterative predictions to improve generation consistency and robustness.
Extensive experiments confirm the effectiveness of DLMRec and demonstrate discrete diffusion language modeling as a compelling alternative to autoregressive generation for recommendation. Our codes are available at ~\url{https://github.com/ChengyiLIU-cs/DLMRec}
\end{abstract}



\received{20 February 2007}
\received[revised]{12 March 2009}
\received[accepted]{5 June 2009}

\maketitle

\input{sections/Introduction}
\input{sections/Preliminary}
\input{sections/Method}
\input{sections/Experiment}

\input{sections/Related_work}

\input{sections/Conclusion}

\bibliographystyle{ACM-Reference-Format}
\bibliography{sample-manuscript}

\end{document}

%% file: sections/Introduction.tex
\section{Introduction}
\label{Introduction}

Recommender Systems (RS) have become essential in the era of information explosion, alleviating information overload by delivering personalized content across domains ranging from e-commerce to social networking ~\cite{fan2019graph,chen2023fairly}.
Collaborative Filtering (CF) has served as a fundamental paradigm in recommender systems, modeling user preferences from historical user-item interactions ~\cite{fan2020graph}. 
Within the traditional discriminative framework, Graph Neural Network (GNN)-based CF approaches have achieved notable success by learning refined representations of users and items to capture high-order collaborative signals from interaction graphs ~\cite{he2020lightgcn,fan2022graph}. 
More recently, generative recommendation has emerged as a promising direction that models the underlying distribution of user behaviors to directly generate preferred items, offering a more flexible paradigm for capturing complex and evolving user preferences~\cite{zhao2024recommender,yang2023generate,liu2025score}. Following this direction, recommenders empowered by large language models (LLMs) have shown particular promise, leveraging their strong generalization ability and generative capacity for user preference modeling ~\cite{wu2024survey,ning2026retrieval,zheng2024adapting}. 

Benefiting from the scaling of model parameters and training corpora, LLMs exhibit strong capabilities in semantic understanding and reasoning over complex information~\cite{achiam2023gpt,touvron2023llama,fan2024survey, ning2026retrieval}. These properties make them particularly suitable for recommendation, where user preferences are often dynamic, context-dependent, and closely tied to rich textual metadata such as profiles, descriptions, and reviews~\cite{wang2025knowledge,huang2026towards, wang2026mixture}. By leveraging such semantic signals, LLM-based recommender systems move beyond conventional interaction-based matching toward a more expressive paradigm for preference modeling, enabling more accurate, explainable, and context-aware recommendations.

\begin{figure}[tp]
\centering
  \includegraphics[width=0.95\textwidth]{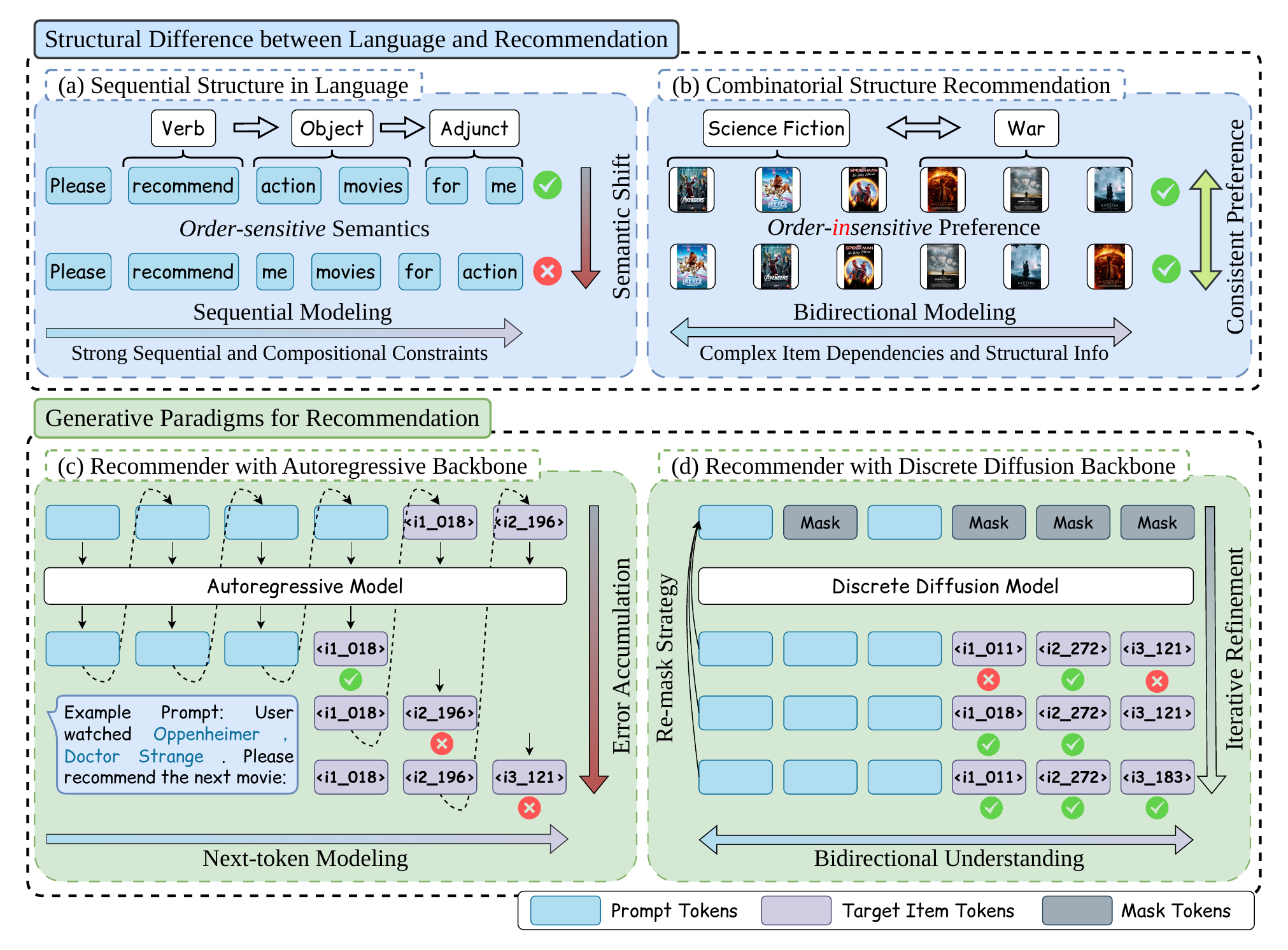}%
\caption{Comparison of structural properties in language and recommendation, and generative paradigms for recommendation. Language follows strong left-to-right sequential dependencies, while recommendations exhibit combinatorial inter-item relationships and order-insensitive preferences. This structural gap suggests the potential of diffusion-based generation for recommendation, as its bidirectional modeling and iterative refinement better capture diverse global user preferences.}
\Description{Comparison of diffusion-based RS across different state-spaces.}
\label{fig:Autoregressive-diffusion} 
\end{figure}

Despite their strong empirical performance, most existing LLM-based generative recommendation methods rely on an autoregressive modeling paradigm, which may be suboptimal for recommendation~\cite{zhu2024collaborative,shi2025llada}.
The limitation mainly stems from the next-token prediction objective, originally designed for language generation, where tokens exhibit strong left-to-right sequential and compositional dependencies ~\cite{zhao2026simgr}, as illustrated in Fig.~\ref{fig:Autoregressive-diffusion}.
In recommendation, however, user preferences are primarily determined by complex relationships among interacted items rather than by their observed ordering. The interaction sequence mainly serves as a behavioral trace, while the observed order is often noisy and influenced by incidental factors~\cite{yang2023debiased}. Consequently, the key challenge lies in modeling inter-item dependencies and structural preference patterns instead of strictly following the interaction order ~\cite{lin2024discrete,xie2024breaking}.
Beyond the objective mismatch, next-token prediction conditions each generation step only on the observed prefix, restricting access to potentially informative future interactions~\cite{mu2026masked,israel2026enabling}. This prefix-constrained modeling paradigm naturally favors local sequential dependencies, while limiting the model's ability to capture global inter-item relationships and structural preference patterns~\cite{zhang2025diffusion,sahoo2024simple}.
Furthermore, autoregressive generation commits to each prediction once it is produced, preventing subsequent corrections and causing early prediction errors to propagate throughout the decoding process~\cite{shi2025llada,li2025breaking}.

\begin{figure}[htbp]
    \centering
    \begin{subfigure}[t]{0.31\textwidth}
        \centering
        \includegraphics[width=\linewidth]{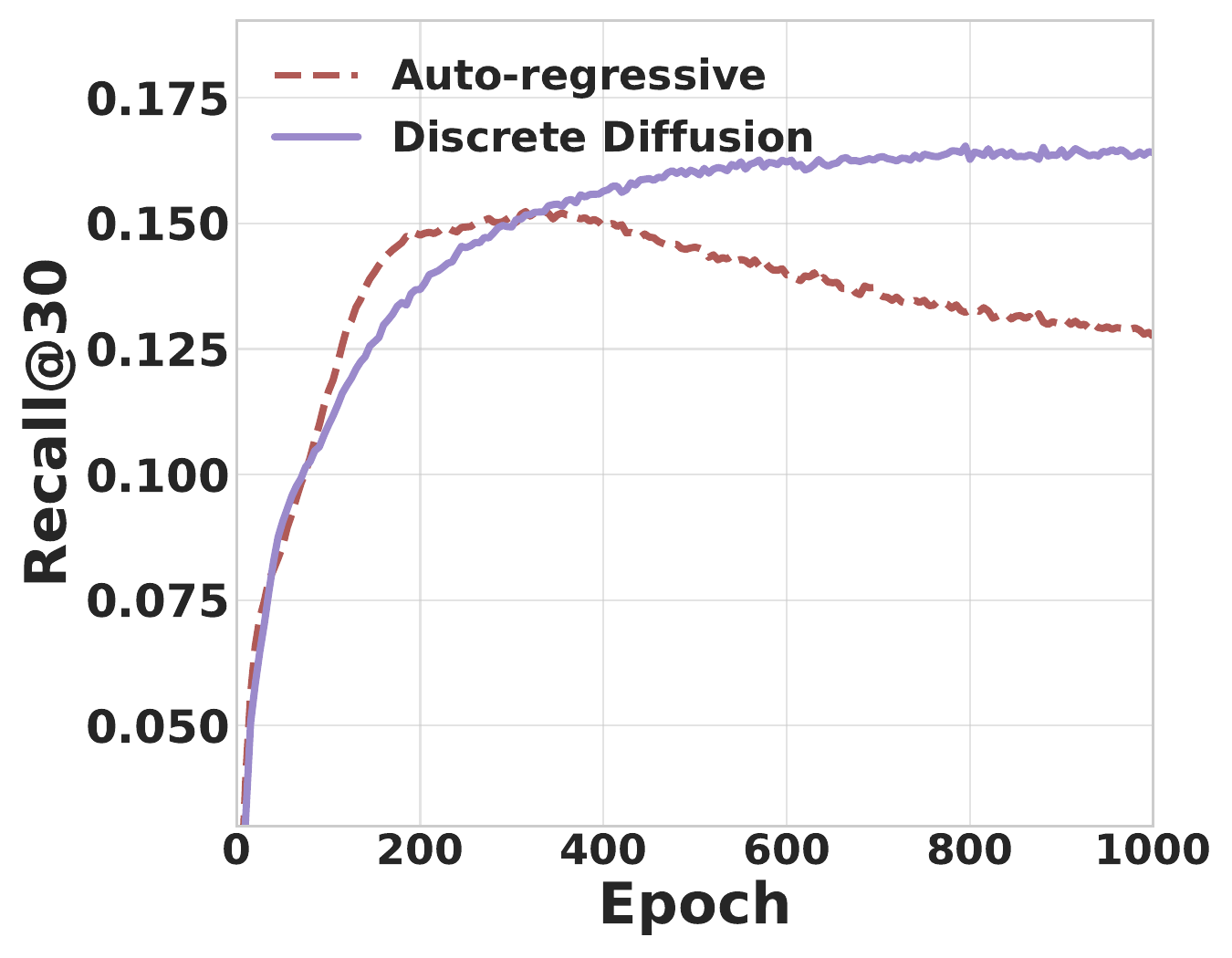}
        \label{fig:pre_curve}
    \end{subfigure}\hfill
    \begin{subfigure}[t]{0.31\textwidth}
        \centering
        \includegraphics[width=\linewidth]{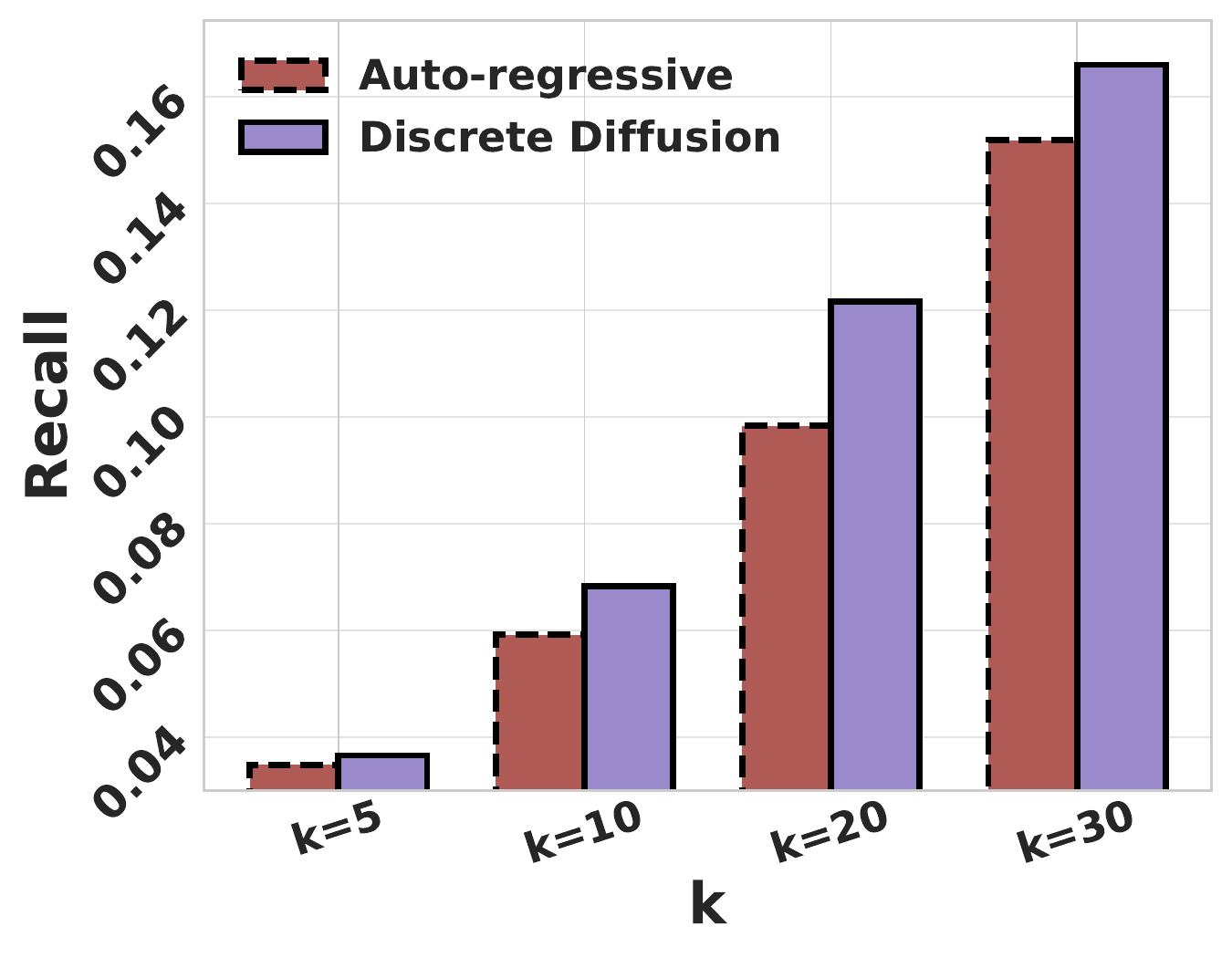}
        \label{fig:pre_recall}
    \end{subfigure}\hfill
    \begin{subfigure}[t]{0.31\textwidth}
        \centering
        \includegraphics[width=\linewidth]{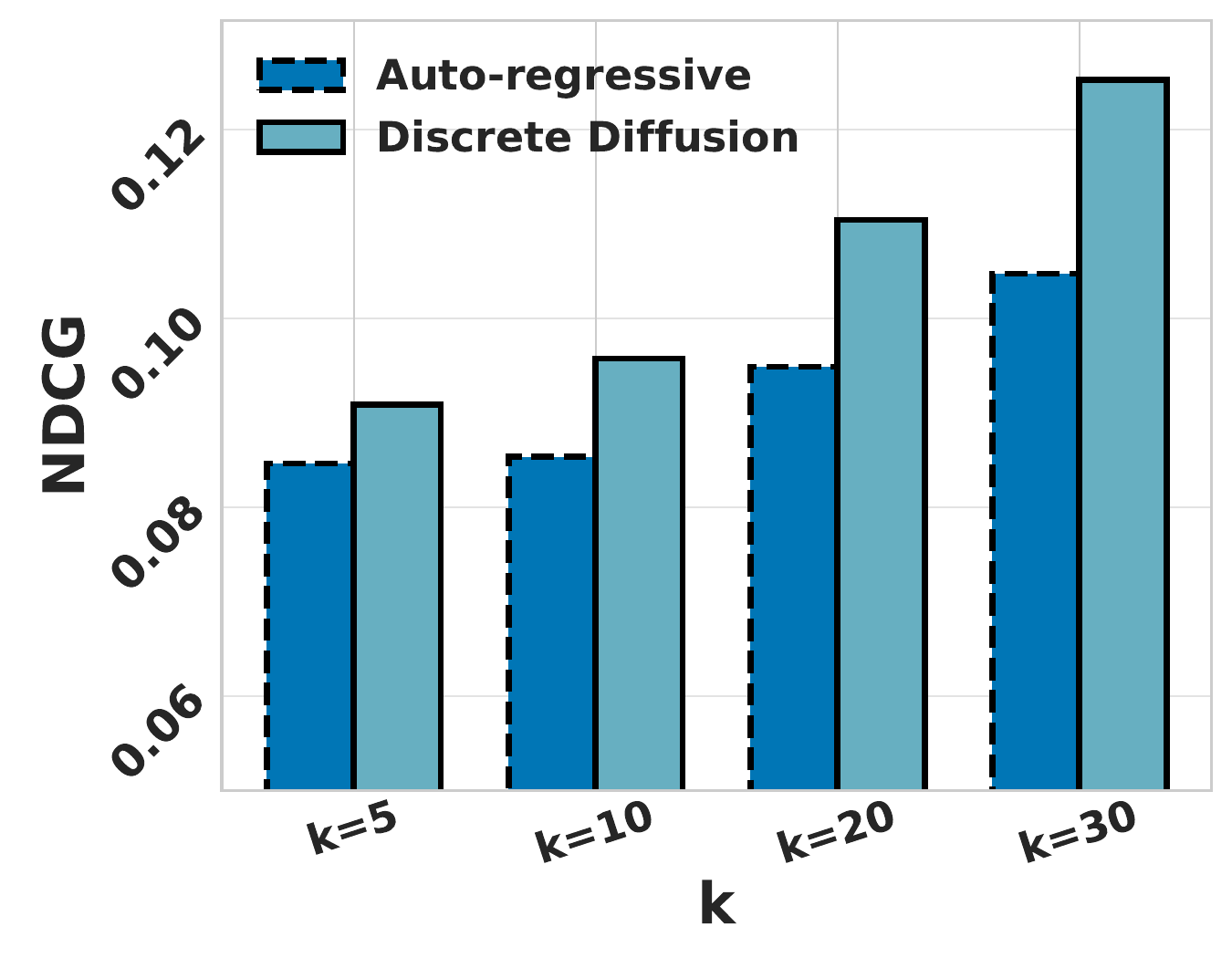}
        \label{fig:pre_ndcg}
    \end{subfigure}
    \caption{Preliminary results on MovieLens-1M comparing autoregressive and discrete diffusion recommendation models. Both models use the same pretrained LightGCN encoder to represent interaction sequences and predict the next item, with identical parameter scales for fair comparison. The key difference lies in the generative paradigm: the autoregressive model performs unidirectional next-item prediction with causal masking, while the diffusion model adopts a bidirectional denoising process based on an absorbing-state diffusion mechanism~\cite{austin2021structured}. Results demonstrate that the diffusion model achieves more stable training dynamics and outperforms the autoregressive baseline in terms of Recall and NDCG, highlighting the potential of diffusion-based generative recommendation.}
    \Description{Preliminary results}
    \label{fig:preliminary_study}
\end{figure}

Recent advances in diffusion language models (DLMs) provide a promising alternative for LLM-based generative recommendation ~\cite{nie2025large, ye2025dream, song2025seed}.
Instead of relying on an autoregressive next-token generation, discrete diffusion language models learn data distributions by progressively masking tokens into an absorbing state and then iteratively recovering them through denoising, yielding a generation process based on bidirectional modeling and whole-sequence refinement ~\cite{austin2021structured}.
Representative studies such as LLaDA suggest that DLMs can achieve capabilities comparable to autoregressive LLMs of similar scale, while also showing promise in learning effectively from limited data, which is particularly appealing for recommendation scenarios with sparse historical interactions ~\cite{nie2024scaling, prabhudesaidiffusion, ni2025diffusion, wang2025diffusion, gao2025mindrec}. 
More importantly, the diffusion objective is naturally aligned with recommendation, which fundamentally requires recovering latent preference structure from sparse, incomplete, and noisy observations rather than merely predicting the next item~\cite{wang2023diffusion,liu2026continuous}.
Such a formulation further provides a flexible framework for modeling diverse interest patterns and exploiting global contextual information more effectively.
To validate this intuition in recommendation scenarios, we conduct a preliminary comparative study on the MovieLens-1M dataset. 
As shown in Fig.~\ref{fig:preliminary_study}, the results are consistent with prior observations in the literature, indicating that discrete diffusion models exhibit stronger potential than autoregressive baselines, with more stable training dynamics and improved performance in terms of Recall and NDCG.

However, directly adapting discrete diffusion language models to recommendation remains challenging.
First, the discrete representations of user-item interactions are required to be both compatible with diffusion language modeling and expressive enough to capture collaborative semantics.  
Existing sequential token decomposition schemes, such as RQ-style decomposition~\cite{rajput2023recommender,deng2025onerec}, introduce inherent sequential dependencies that are misaligned with the parallel denoising nature of diffusion models. On the other hand, whole-item representations are often overly coarse, making it difficult to preserve fine-grained collaborative signals while maintaining unique and conflict-free token assignments~\cite{qu2025tokenrec,takase2020all}.
Second, the standard random masking strategy falls short of meeting the task-specific requisites of recommendation, as it overlooks the fundamental divergence in optimization between language generation and preference modeling. While diffusion language models are designed to optimize linguistic coherence, recommendations aim to learn latent user preferences from complex high-order item interactions. In addition, recommendation requires grounding newly introduced user and item tokens into a shared collaborative semantic space, a process that is disrupted by indiscriminate corruption. As a result, treating all tokens uniformly weakens collaborative signals and undermines the global coherence required for effective preference modeling.
Third, standard remasking-based diffusion decoding repeatedly discards intermediate predictions during iterative refinement, leading to suboptimal utilization of previously inferred signals and potentially weakening global coherence in recommendation generation~\cite{ye2026rejection,zheng2025continuously,pynadath2025candi}.

To overcome the aforementioned challenges, we propose a novel \underline{D}iffusion \underline{L}anguage \underline{M}odel \underline{Rec}ommendation framework, DLMRec, to address the above challenges. 
Specifically, DLMRec proposes a collaborative-aware stochastic tokenizer (CAST) that transforms multi-hop collaborative signals from the user-item graph into discrete token representations. Within this module, collaborative semantics from different propagation depths are explicitly disentangled into hop-wise semantic tokens, while a topology-stochastic assignment mechanism is introduced to handle the inherent ambiguity of graph-based discretization and ensure compatibility with the diffusion process.
To bridge the gap between language modeling objectives and recommendation-specific semantics, we propose a curriculum-driven training strategy that progressively aligns user preference learning at both item and token levels. To further enhance this alignment, we incorporate negative samples into the training objective, encouraging the model to distinguish positive interactions from plausible distractors, thereby improving its discriminative ability in capturing fine-grained collaborative patterns.
Finally, we introduce a stability-aware voting mechanism for final prediction, which leverages the full refinement trajectory to produce the final output. By aggregating predictive evidence across all iterative steps, this mechanism selectively preserves confident predictions while resolving instability in uncertain positions, resulting in more consistent and robust recommendations.

This paper makes the following main contributions:
\begin{itemize}
    \item We propose DLMRec, a novel discrete diffusion language model tailored for recommendation. Unlike autoregressive paradigms that generate items sequentially in a left-to-right manner, DLMRec leverages bidirectional context modeling and iterative refinement to progressively improve predictions toward global coherence, establishing a diffusion-based paradigm for LLM-driven recommendation.

    \item We introduce three key designs to adapt DLMs to recommendation: a collaborative-aware stochastic tokenizer, a curriculum-driven training strategy, and a voting-based iterative refinement mechanism.
    The stochastic tokenizer constructs discrete interaction representations that preserve multi-hop collaborative semantics while ensuring diffusion compatibility; the curriculum training strategy progressively aligns the denoising process with preference recovery; and the voting-based refinement mechanism stabilizes generation by aggregating predictions across iterative steps, thereby enhancing the coherence of final recommendations.

    \item We conduct extensive experiments to evaluate DLMRec. The results validate its effectiveness and underscore the promise of discrete diffusion language modeling as a compelling alternative to autoregressive generation for recommendation.
\end{itemize}

%% file: sections/Preliminary.tex
\section{Preliminary}
\label{Preliminary}

In this section, we introduce the key notations and definitions used throughout this paper, followed by a brief overview of vector quantization (VQ) techniques and discrete diffusion algorithms.

\subsection{Notations and Definitions}

The primary objective of recommender systems is to infer user preferences by analyzing historical behavioral data, including clicks, browsing activities, and transaction histories. We define $\mathcal{U}={u_1, \dots ,u_M}$ as the user set containing $M$ users, and $\mathcal{V}={v_1, \dots ,v_N}$ as the item set comprising $N$ items. The interaction history for any given user $u_m \in \mathcal{U}$ is formulated as a sequence $\mathbf{R}_{u_m} = [\mathbf{v}_1, \mathbf{v}_2, \dots, \mathbf{v}_l]$, where $l$ represents the predetermined length of the interaction sequence.
To model collaborative filtering signals effectively, both users and items are projected into compact latent representations. More precisely, the user embedding matrix is denoted as $\mathbf{P} = [\mathbf{p}_1, \dots, \mathbf{p}_M]^T \in \mathbb{R}^{M \times d}$, and the item embedding matrix is expressed as $\mathbf{Q} = [\mathbf{q}_1, \dots, \mathbf{q}_N]^T \in \mathbb{R}^{N \times d}$, where $d$ specifies the dimensionality of the latent space.

In the context of LLM-based recommender systems, recommendation tasks are reformulated within a language-centric framework. User interaction histories are converted into textual prompts that are subsequently fed into large language models for processing.
More formally, each user $u_m$ and item $v_n$ is associated with corresponding textual prompts $\mathcal{P}_{u_m}$ and $\mathcal{P}_{v_n}$, respectively. Building upon this notation, the input to an LLM-based recommender and its corresponding output can be expressed as:
\begin{equation}
    \mathcal{Y}i = \text{DLM}(\mathbf{P} (\mathcal{P}_{u_m}, {\mathcal{P}_{v_n} \mid v_n \in \mathbf{R}_{u_m}})),
\end{equation}
where $\mathbf{P}(\cdot)$ denotes the prompt assembly function that constructs a unified textual sequence containing user tokens and historical item tokens.
The output $\mathcal{Y}_i$ generated by the DLM can take various forms, such as item identifiers, special tokens, or dense embeddings, depending on the underlying model architecture and task formulation.

\subsection{User and Item Tokenization}

Large language model-based recommender systems typically rely on discrete representations of users and items to align with the linguistic structure of natural language. While the straightforward ID-based paradigm assigns unique identifiers to each user and item, it is limited in both semantic expressiveness and scalability, as vocabulary size grows linearly with the number of items.
Moving beyond simple indexing, vector quantization has emerged as the predominant tokenization strategy in LLM-based recommender systems~\cite{hua2023index}. This approach projects multifaceted user or item attributes into a finite set of discrete semantic tokens. Specifically, collaborative signals from user-item interaction graphs (via GNN-based encoders) or semantic information from textual descriptions (via pre-trained language encoders) are transformed into compact discrete representations. A vector-quantized tokenizer is typically achieved with an encoder-decoder architecture, formalized as follows:
\begin{align}
    \boldsymbol{z} &= f_{\mathrm{enc}}(\mathbf{e}), \\
    s &= \arg\min_{k} \left\| \boldsymbol{z} - \mathbf{c}_s \right\|_2^2, \\
    \hat{\mathbf{e}} &= f_{\mathrm{dec}}(\mathbf{c}_{s}),
\end{align}
where $\mathbf{e}$ denotes the feature representation of a user or item, $\mathbf{c}_s$ represents the $s$-th codeword selected from the codebook $\mathcal{C} \in\mathbb{R}^{L\times d_c}$, and $\hat{\mathbf{e}}$ denotes the reconstructed representation decoded from the quantized codeword.
Under this scheme, a user $u_m$ or an item $v_n$ is tokenized by the index $s$ of its nearest codeword and by the corresponding codeword embedding $\mathbf{c}_s$. This yields discrete representations that preserve semantic information while avoiding linear vocabulary growth.

\subsection{Discrete Diffusion Language Model}

Recent advances in discrete diffusion models have shown promise for text generation, with LLaDA being a notable example. Unlike continuous diffusion models that corrupt data with Gaussian noise, discrete diffusion approaches apply state transformations, typically token masking, to progressively degrade text sequences until they reach a stationary distribution, known as the absorbing state.
Formally, for an arbitrary token $\mathbf{x}$ in the text sequence, the forward diffusion process at timestep $t$ can be described as a Markov transition:
\begin{equation}
    q(\mathbf{x}_t \mid \mathbf{x}_{t-1})
    = \mathrm{Cat}\!\left(\mathbf{x}_t;\, \mathbf{p} = \mathbf{x}_{t-1}\mathbf{Q}_t\right),\
\end{equation}
where $\mathrm{Cat}(\cdot)$ denotes a categorical distribution, and $\mathbf{Q}_t \in \mathbb{R}^{V \times V}$ represents the transition matrix applied element-wise across sequence positions, with $V$ being the vocabulary size.
The transition matrix $\mathbf{Q}_t$ is specifically designed to model the masking operation. Each entry $[\mathbf{Q}_t]_{i,j}$ at diffusion step $t$ is defined as:
\begin{equation}
    [\mathbf{Q}_t]_{i,j} =
    \begin{cases}
    1, & \text{if } i = j = [\mathrm{M}], \\[4pt]
    \beta_t, & \text{if } j = [\mathrm{M}],\, i \neq [\mathrm{M}], \\[4pt]
    1 - \beta_t, & \text{if } i = j \neq [\mathrm{M}].
    \end{cases}
\end{equation}
where $[\mathrm{M}]$ denotes the mask token index and $\beta_t \in [0,1]$ controls the masking rate at step $t$.
By applying the transition from the initial state $\mathbf{x}_0$, the marginal distribution of a token at timestep $t$ can be derived as:
\begin{equation}
    q(x_t^i \mid x_0^i) =
    \begin{cases}
    \bar{\alpha}_t, & \text{if } x_t^i = x_0^i, \\[4pt]
    1 - \bar{\alpha}_t, & \text{if } x_t^i = [\mathrm{M}],
    \end{cases}
    \label{eq:diff_forward}
\end{equation}
where $\bar{\alpha}_t = \prod_{i=1}^{t}(1-\beta_i)$ represents the cumulative probability that the token remains unmasked up to step $t$.
The central architecture of discrete diffusion comprises a parametric prediction model $p_\theta(\cdot \mid \mathbf{x}_t)$ that performs parallel inference over all masked tokens given the noised input $\mathbf{x}_t$. The model parameters $\theta$ are learned by minimizing a cross-entropy objective that is evaluated exclusively on masked token positions:
\begin{equation}
    \mathcal{L}(\theta)\triangleq-\mathbb{E}_{t, x_0, x_t}\left[\frac{1}{t}\sum_{i=1}^{L}\mathbf{1}[x_t^i = \mathrm{[M]}]\log p_{\theta}(x_0^i \mid x_t)\right],
    \label{eq:loss_diffusion_pre}
\end{equation}
where $\mathbf{x}_0$ denotes a training sample from the data distribution, $t$ is a timestep uniformly sampled from $0$ to $T$, where $T$ is the maximum diffusion steps, and $\mathbf{x}_t$ is the noised sequence obtained via the forward diffusion process defined in Eq.~\ref{eq:diff_forward}. The sequence length is denoted by $L$, and the indicator function $\mathbf{1}[\cdot]$ ensures that the loss is computed only over positions where tokens have been masked.
During the inference phase, various remasking strategies are applied to iteratively refine the generation quality.

%% file: sections/Method.tex
\section{The Proposed Method}
\label{Method}
In this section, we present an overview of the proposed framework, followed by a detailed description of its key components.

\begin{figure}[t!]
\centering
  \includegraphics[width=1.0\textwidth]{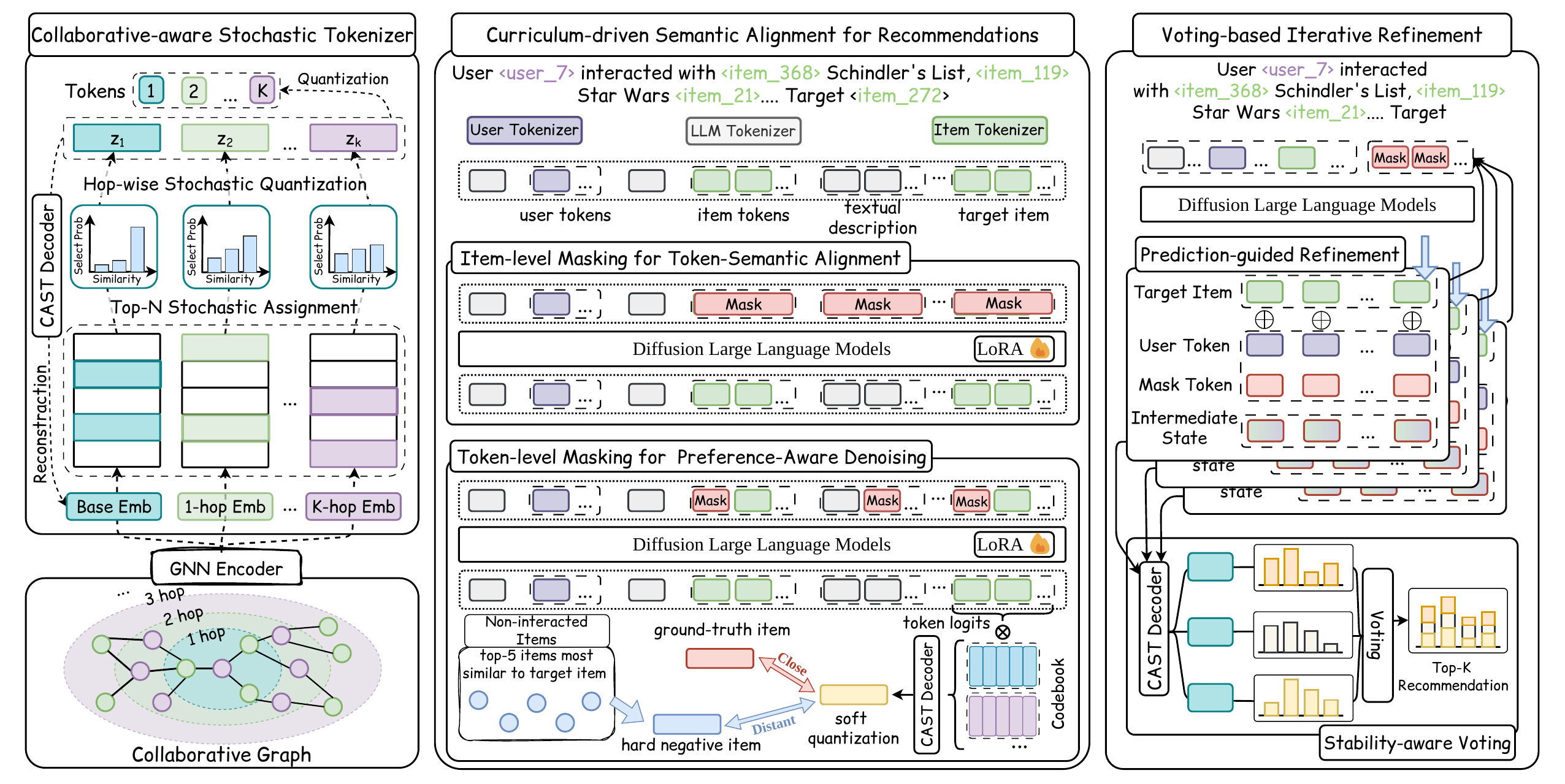}%
\caption{The proposed DLMRec framework contains three main modules: collaborative-aware stochastic tokenizer,  curriculum-driven training strategy, and stability-aware voting mechanism.}
\Description{DLMRec framework}
\label{fig:DLMRec} 
\end{figure}

\subsection{An Overview of the Proposed Framework}
We propose a discrete diffusion language modeling framework for recommendation, as illustrated in Fig.~\ref{fig:DLMRec}. Initially, DLMRec encodes each user and their historical item interactions into discrete token representations that preserve collaborative semantics while ensuring compatibility with the diffusion process. A hybrid prompt is then constructed by augmenting these user and item tokens with textual information to enrich preference modeling. Next, we design a curriculum-driven training strategy that aligns diffusion-based denoising with preference learning. The model is trained to recover progressively masked item tokens under varying corruption levels, capturing both item-level semantics and high-order collaborative dependencies, while negative samples are incorporated to sharpen discriminative capability. Finally, during inference, DLMRec performs iterative refinement over the target item tokens. A stability-aware decoding mechanism distinguishes confident predictions from uncertain ones: stable tokens are committed early, while unstable tokens are refined across multiple diffusion steps. A voting-based aggregation strategy then synthesizes the full refinement trajectory to produce the final recommendation.

\subsection{Collaborative-Aware Stochastic Tokenizer}

To bridge collaborative recommendation and diffusion language modeling, it is necessary to construct discrete representations that are compatible with the diffusion process while retaining recommendation-specific semantics. The ideal representation should preserve collaborative similarity structures and assign each user and item unique, non-conflicting discrete codes.
However, existing tokenization approaches present two key challenges. First, hierarchical tokenization methods (e.g., RQ-style) impose sequential dependencies that conflict with the bidirectional denoising of DLMs ~\cite{shi2025llada,rajput2023recommender,wang2024learnable}. Second, text-derived representations often fail to capture collaborative semantics, as pre-trained encoders are optimized for linguistic modeling rather than recommendation objectives~\cite{zheng2024adapting,liao2024llara}.
To address these issues, we propose a collaborative-aware stochastic tokenizer (CAST), which learns discrete representations from collaborative signals in the user–item interaction graph. Specifically, CAST first extracts multi-hop collaborative embeddings to encode high-order interaction structure, and then performs stochastic quantization to map them into a discrete token space.
In this way, the tokenized representations serve as discrete semantic units for subsequent preference modeling. As the tokenization procedures for users and items are symmetric, we detail the user-side formulation below, and the item-side construction follows an analogous logic.


\subsubsection{Collaborative Semantic Representation}

In collaborative recommendation, user and item representations are shaped not only by individual attributes but, more importantly, by the multi-hop relational structure of the interaction graph, making graph neural networks (GNNs) a natural backbone for modeling collaborative signals~\cite{he2020lightgcn,wang2019neural}. Representations at different propagation depths encode distinct levels of preference information, ranging from direct interactions to higher-order collaborative dependencies. Accordingly, our tokenizer encodes these multi-hop signals into discrete tokens, where each token corresponds to a specific propagation depth to preserve structured collaborative information. Since these tokens are constructed independently from multi-hop representations, they do not impose explicit inter-token dependencies and are naturally compatible with the bidirectional denoising process of DLMs.


To instantiate this architecture, we maintain a set of codebook families for users, $\mathcal{C}=\{\mathbf{C}^{1},\mathbf{C}^{2},\ldots,\mathbf{C}^{K}\}$, where each $\mathbf{C}^{k}$ denotes the $k$-th sub-codebook containing $L$ learnable codewords of dimension \(d_c\). Different codebooks are responsible for discretizing collaborative semantics from different propagation depths, enabling each token position to capture a distinct level of preference abstraction.

To obtain such hop-wise collaborative semantics, we adopt LightGCN~\cite{he2020lightgcn} as the backbone encoder over the historical interaction graph. At the $l$-th propagation layer, LightGCN updates representations of user through neighborhood aggregation:
\begin{equation}
\mathbf{e}^{u,(l)}_m
=
\sum_{v_n \in \mathcal{N}(u_m)}
\frac{1}
{\sqrt{|\mathcal{N}(u_m)|\,|\mathcal{N}(v_n)|}}
\mathbf{e}^{v,(l-1)}_{v_n},
\quad
\text{with }
\mathbf{e}^{u,(0)}_m = \mathbf{p}_m,
\end{equation}
where \(\mathcal{N}(u_m)\) and \(\mathcal{N}(v_n)\) denote the neighbor sets of user \(u_m\) and item \(v_n\), respectively. 
Through iterative propagation, \(\mathbf{e}^{(l)}\) captures collaborative signals from the \(l\)-hop neighborhood, with shallow layers focusing on direct interaction patterns and deeper layers encoding higher-order collaborative relations. 
While we use LightGCN as the backbone, our tokenization mechanism generalizes to any GNN-based collaborative encoder producing multi-hop representations.
Instead of aggregating all layers into a single representation, we explicitly preserve the hop-wise user embeddings
\begin{equation}
\mathbf{H}^{u}_m = \{\mathbf{e}^{u,(0)}_m, \mathbf{e}^{u,(1)}_m, \ldots, \mathbf{e}^{u,(K-1)}_m\},
\end{equation}
where each $\mathbf{e}^{u,(k-1)}_m$ is associated with the \(k\)-th token position, yielding a token sequence that captures multi-granularity collaborative semantics without imposing sequential dependencies.

\subsubsection{Hop-wise Stochastic Quantization}
Given the hop-wise collaborative representations, we next discretize them into semantic tokens. While deterministic nearest-code assignment provides a simple quantization strategy, it may be overly restrictive for collaborative recommendation. Because user preferences are inherently diverse and stochastic, graph-based representations often encode multiple collaborative signals and cannot always be faithfully associated with a single codeword ~\cite{xie2024breaking, yuan2026graver}. Consequently, hard assignment imposes sharp semantic boundaries and discards valuable collaborative uncertainty~\cite{zhao2026mitigating}.
Motivated by these observations, we propose a stochastic quantization scheme that samples codewords from a similarity-aware assignment distribution, thereby preserving uncertainty-aware semantic structure.

Specifically, we first compute the cosine similarity between each hop-wise user representation $\mathbf{e}^{(k-1)}_m$ and the codewords $\mathbf{c}_j^{(k)}$ in the corresponding sub-codebook $\mathbf{C}^{(k)}$:
\begin{equation}
s_{m,j}^{(k)}
=   
\frac{\langle \mathbf{e}^{(k-1)}_m, \mathbf{c}_j^{(k)} \rangle}
{\|\mathbf{e}^{(k-1)}_m\|\,\|\mathbf{c}_j^{(k)}\|},
\qquad j=1,\ldots,L.
\label{eq:similarity}
\end{equation}
A direct softmax over all codewords may assign non-negligible probability mass to many low-similarity candidates, thereby introducing noisy token assignments. To alleviate this issue, we restrict the assignment space to the top-$\mathcal{S}$ most similar codewords and normalize the assignment probabilities only within this subset.
The stochastic transition probability for $\mathbf{e}^{(k-1)}_m$ to the $j$-th codeword is defined as
\begin{equation}
\begin{aligned}
P(w_m^{k} = j)
&=
\begin{cases}
\dfrac{\exp\!\left(s_{m,j}^{(k)}/\tau_k\right)}
{\sum\limits_{r \in \mathrm{Top}\mathcal{S}(\mathbf{e}_m^{(k-1)}, \mathbf{C}^{(k)})}
\exp\!\left(s_{m,r}^{(k)}/\tau_k\right)},
& \text{if } j \in \mathrm{Top}\mathcal{S}(\mathbf{e}_m^{(k-1)}, \mathbf{C}^{(k)}), \\[8pt]
0, & \text{otherwise},
\end{cases}
\\[4pt]
&\hspace{3.6em}
\tau_k
=
\tau_{\min}+(\tau_{\max}-\tau_{\min})\left(\frac{k}{K}\right)^{\alpha},
\end{aligned}
\label{eq:transition_distribution}
\end{equation}
where \(\tau_k\) is a hop-aware temperature that increases monotonically with the hop index to control the sharpness of the assignment distribution, $\tau_{\max}$ and $\tau_{\min}$ are its lower and upper bounds, and $\alpha$ controls the growth rate. 
Intuitively, lower-order hops adopt smaller temperatures for sharper assignments to capture direct and stable preference signals, while higher-order hops use larger temperatures to enable softer, more stochastic assignments for abstract and uncertain collaborative semantics. In this way, the degree of quantization stochasticity is aligned with the semantic uncertainty across different propagation depths.

\subsubsection{Semantic Reconstruction Decoder}

Following stochastic quantization, the user $u_m$ is assigned with the discrete tokens $\mathcal{W}_{u_m}=\{w^{1}_m, w^{2}_m, \ldots, w^{K}_m\}$, together with their corresponding token embeddings $[\mathbf{c}^{1}_{w^{1}_m}, \mathbf{c}^{2}_{w^{2}_m} \ldots,  \mathbf{c}^{K}_{w^{K}_m}]$.
To enable reconstruction, we concatenate these token embeddings and apply a decoder:
\begin{equation}
\hat{\mathbf{e}}_m = f_{dec}([\mathbf{c}^{1}_{w^{1}_m}, \mathbf{c}^{2}_{w^{2}_m}, \ldots ,\mathbf{c}^{K}_{w^{K}_m}]),
\end{equation}
where $\hat{\mathbf{e}}_m$ denotes the reconstructed embedding of user $u_m$.
The same quantization and reconstruction procedure is applied to items in an analogous manner as demonstrated in Algorithm \ref{alg:cast}.

\begin{algorithm}[t]
\caption{Collaborative-aware Stochastic Tokenizer}
\label{alg:cast}
\begin{algorithmic}[1]

\REQUIRE User-item graph $\mathcal{G}$, pretrained graph encoder $f_{\mathrm{enc}}$, user/item codebook families $\mathcal{C}^{u}, \mathcal{C}^{i}$, candidate size $\mathcal{S}$, temperature bounds $\tau_{\min}$ and $\tau_{\max}$, growth factor $\alpha$, decoder $f_{\mathrm{dec}}$, target user/item $(u_m \text{ or } v_n)$
\ENSURE Discrete token sequence $\mathbf{w} = [w^{1}, \ldots, w^{K}]$ and reconstructed embedding $\hat{\mathbf{e}}$ 

\STATE Obtain hop-wise collaborative representations of user $u_m$ or item $v_n$ from the pretrained graph encoder $f_{\mathrm{enc}}$, written as $\mathcal{H}=\{\mathbf{e}^{(0)}, \mathbf{e}^{(1)}, \ldots, \mathbf{e}^{(K-1)}\}$.
\FOR{$k=1$ to $K$}
    \STATE Compute cosine similarities between $\mathbf{e}^{(k-1)}$ and each codeword in the corresponding codebook via Eq. (\ref{eq:similarity})
    \STATE Obtain the top-$\mathcal{S}$ most similar codewords:
    \STATE Compute the stochastic assignment distribution over Eq. (\ref{eq:transition_distribution})
    \STATE Sample the token index $w^{k} \sim P(w^{k})$
    \STATE Retrieve the corresponding token embedding $\mathbf{c}_{w^{k}}^{(k)}$
\ENDFOR

\STATE Form the discrete token sequence:
\[
\mathbf{\mathcal{W}} = [w^{1}, w^{2}, \ldots, w^{K}]
\]

\STATE Concatenate token embeddings:
\[
\bar{\mathbf{c}} = [\mathbf{c}_{w^{1}}^{(1)}, \mathbf{c}_{w^{2}}^{(2)}, \ldots  ,\mathbf{c}_{w^{K}}^{(K)}]
\]

\STATE Reconstruct the original embedding through the decoder:
\[
\hat{\mathbf{e}} = f_{\mathrm{dec}}(\bar{\mathbf{c}})
\]

\RETURN $\mathbf{w},\hat{\mathbf{e}}$
\end{algorithmic}
\end{algorithm}

\subsubsection{Training Objective}

The tokenizer and decoder are jointly trained to preserve collaborative semantics while learning stable discrete tokens. We employ a reconstruction loss to ensure the decoder $f_{\text{dec}}$ faithfully recovers the original user representation $\mathbf{p}_m$:
\begin{equation}
    \mathcal{L}{\text{rec}} = |\mathbf{p}_m - \hat{\mathbf{e}}_m|^{2}.
\end{equation}
However, reconstruction alone is insufficient to guarantee stable quantization. We therefore leverage a codebook loss and a commitment loss to regulate the alignment between collaborative representations and their assigned codewords:
\begin{equation}
\begin{aligned}
    \mathcal{L}_{\text{cb}} &= \|\text{sg}[\mathbf{e}^{(k-1)}_m] - \mathbf{c}^{k}_{w^{k}_m}\|^2, \\
    \mathcal{L}_{\text{commit}} &= \|\mathbf{e}^{(k-1)}_m - \text{sg}[\mathbf{c}^{k}_{w^{k}_m}]\|^2, \\
\end{aligned}
\end{equation}
where $\mathrm{sg}[\cdot]$ denotes the stop-gradient operator. In particular, $\mathcal{L}_{\text{cb}}$ updates the codebook embeddings toward the collaborative representations, while $\mathcal{L}_{\text{commit}}$ encourages the representations to stay close to their assigned codewords.

While the reconstruction loss enforces pointwise consistency, it does not explicitly promote discriminative alignment in the embedding space. Since the downstream recommendation task is inherently retrieval-oriented, we introduce a contrastive loss to align each reconstructed representation with its original embedding while separating it from other users in the batch. Formally, for a mini-batch of size \(B\), the contrastive loss is defined as
\begin{equation}
\mathcal{L}_{\mathrm{cl}}
=
-\frac{1}{B}\sum_{m=1}^{B}
\log
\frac{\exp\!\left(\mathrm{sim}(\hat{\mathbf{e}}_m,\mathbf{p}_m)/\tau\right)}
{\sum_{n=1, n \neq m}^{B}\exp\!\left(\mathrm{sim}(\hat{\mathbf{e}}_m,\mathbf{p}_n)/\tau\right)},
\end{equation}
where \(\hat{\mathbf{e}}_m\) is the reconstructed embedding, \(\mathbf{p}_m\) is the corresponding original embedding, and the remaining \(\mathbf{p}_n\) in the batch serve as negative samples.
The final training objective is defined as
\begin{equation}
\mathcal{L} = \mathcal{L}{\text{rec}} + \mathcal{L}_{\text{cb}} + \lambda_{cm}\mathcal{L}_{\text{commit}} + \lambda_{cl}\mathcal{L}_{\mathrm{cl}},
\end{equation}
where $\lambda_{cm}$ and $\lambda_{cl}$ are balancing coefficients.
For clarity, the above formulation is presented for user representations. The item-side training procedure follows the same formulation analogously.

\subsection{Curriculum-driven Semantic Alignment for Recommendations}

After obtaining discretized user and item tokens, a key challenge is how to integrate these out-of-vocabulary recommendation tokens into the semantic space of the diffusion language model. Since DLMs are primarily trained on natural language corpora, adapting them to understand recommendation-derived tokens poses a significant challenge. To address this, we propose a curriculum-driven semantic alignment strategy that progressively adapts the DLM from semantic grounding to task-specific recommendation modeling.
First, we conduct item-level alignment to ground item tokens in the semantic space of the model and establish stable item representations. Then, we perform recommendation-oriented adaptation at the token level, enabling the model to capture user preference patterns and generate recommendation-tailored outputs. Through this progressive curriculum, the DLM achieves unified alignment of item semantics and recommendation objectives.

\subsubsection{Hybrid Recommendation Prompt Construction}

To adapt diffusion language models to recommendation, we construct a hybrid prompt that unifies collaborative signals with semantic item descriptions. The quantization module abstracts user-item interaction patterns and graph-topological structures into discrete collaborative tokens, while textual descriptions offer complementary semantic signals. By integrating these two modalities, the prompt provides the model with a holistic view that jointly captures collaborative preferences and semantic understanding during generation.

\begin{figure}[htp]
\centering
  \includegraphics[width=0.8\textwidth]{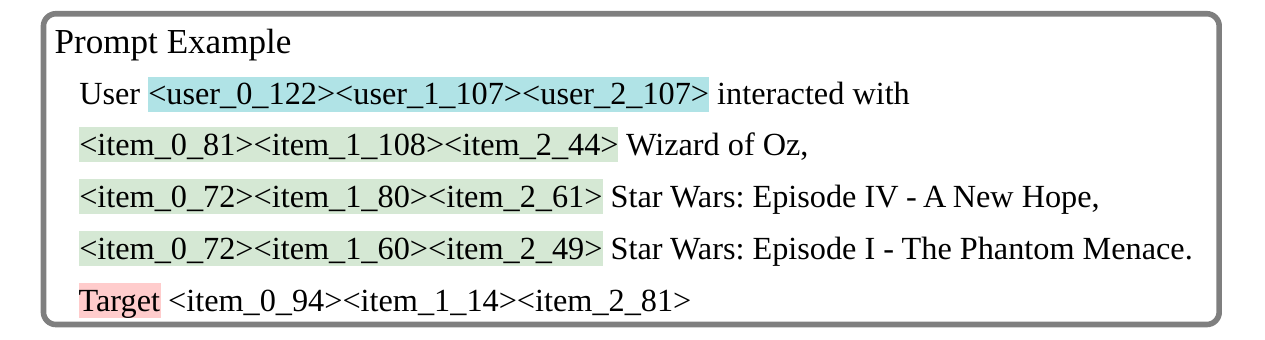}
\caption{Illustration of a hybrid recommendation prompt with a real example from the MovieLens-1M dataset. In the quantization process, the number of codebooks is set to $K=3$, and each codebook contains $L=128$ codewords. Each item is represented by multiple discrete tokens together with its textual title. For example, $<\text{item}\_0\_72>$ denotes that the corresponding item token is selected as the 72nd codeword from the first codebook.}
\Description{Prompt}
\label{fig:prompt} 
\end{figure}

For each user, the prompt comprises a tokenized user identifier, a sequence of historical interactions, and a target item. Historical items are each represented as a collaborative token paired with its textual description, enabling the model to ground collaborative patterns in semantic context. The target item is represented only by its collaborative token sequence and serves as the prediction target.
Fig.~\ref{fig:prompt} presents a representative example of the hybrid prompt, where textual descriptions are instantiated as movie titles.

\subsubsection{Item-level Masking for Token-Semantic Alignment}

The core of Stage 1 lies in an item-level masking strategy tailored to the hybrid recommendation prompt. Instead of masking tokens independently, we preserve the structural integrity of item-related information during corruption.
Specifically, for each historical item $v_n$, we consider two item-specific blocks in the prompt: the discrete item-token block $\mathcal{W}_{v_n}$ (from CAST), and the textual description block ${\mathcal{P}_{v_n}}$ (from the semantic description). These two blocks are masked at the item level with distinct probabilities. The target item tokens are fully masked, requiring the model to infer the complete item identity from the remaining context. In contrast, the natural-language template tokens remain always visible.
This design encourages the model to recover item semantics from partially observed contexts rather than relying on local token co-occurrence alone, thereby establishing the correspondence between collaborative tokens and their textual descriptions.

For stage 1, we design an easy-to-hard item-level masking strategy that progressively increases corruption difficulty as training proceeds.
In early epochs, the model predominantly encounters lightly perturbed item contexts, facilitating stable learning of token-text correspondences. As training advances, the probability of heavily masked contexts gradually increases, exposing the model to progressively harder denoising challenges. This curriculum design promotes steady alignment of item-level semantics while preventing instability caused by aggressive corruption in the initial phase. Specifically, for each item-level masking unit, the masking probability is defined as:
\begin{equation}
1-\bar{\alpha}_t = \frac{t}{T} +
\lambda \sin\!\left(\frac{t\pi}{T}\right)\left(\frac{n}{\mathcal{N}}-\frac{1}{2}\right),
\label{eq:item_level_mask_prob}
\end{equation}
where $n$ denotes the current training epoch in Stage 1, $\mathcal{N}$ is the total number of Stage 1 epochs, $t$ is the sampled diffusion timestep, $T$ is the maximum diffusion timestep, and $\lambda$ controls the curriculum strength. The resulting value is clipped to $[0,1]$ to determine whether each historical item block is masked. 
To facilitate training, we assign separate curriculum coefficients $\lambda$ to the item-token block and the textual description block, reflecting their different roles in semantic grounding. 
Crucially, jointly masking these two blocks prevents the model from learning a trivial mapping from item names to codebook tokens~\cite{zheng2024adapting, qi2025comp}. Instead, it must infer item semantics from incomplete collaborative and textual contexts, establishing a robust alignment between discrete tokens and their underlying meanings in the semantic space.
Formally, Stage 1 adopts the standard masked reconstruction objective of diffusion language models, denoted as $\mathcal{L}_{diff}$ in Eq.~\ref{eq:loss_diffusion_pre}.

\subsubsection{Token-level Masking for Preference-Aware Denoising}

Stage 2 shifts from item-level semantic alignment to recommendation-oriented preference prediction through token-level masking. Moving to finer-grained token-level denoising encourages the model to capture dependencies both within item tuples and across interaction sequences. To further enhance discriminative capability, we introduce an embedding-space preference loss that pulls predicted representations toward ground-truth targets while pushing them away from hard negatives. Together with the token reconstruction objective, this auxiliary loss strengthens both token-level prediction and recommendation-specific discrimination.

Each token in both target and historical items is independently corrupted under a linear masking schedule. Specifically, given a sampled diffusion step $t \sim \text{Uniform}(0,T)$, the forward corruption process is:
\begin{equation}
    q(x_t^i \mid x_0^i) =
    \begin{cases}
    1-\frac{t}{T}, & \text{if } x_t^i = x_0^i, \\[4pt]
    t, & \text{if } x_t^i = [\mathrm{M}],
    \end{cases}
    \label{eq:linear_diff_forward}
\end{equation}
where each token remains unchanged with probability $1-\frac{t}{T}$ or is replaced by the mask token $[\mathrm{M}]$ with probability $\frac{t}{T}$. Textual descriptions, user tokens, and template text remain unmasked. To ensure meaningful prediction, we enforce that at least one target token is masked at each step.

\begin{algorithm}[tp]
\caption{Curriculum-driven Semantic Alignment for Recommendations}
\label{alg:two_stage_training}
\begin{algorithmic}[1]
\REQUIRE Training set $\mathcal{D}$, pretrained diffusion language model $\textit{DLM}$, frozen CAST tokenizers, Stage 1 epochs $\mathcal{N}$, Stage 2 epochs $\mathcal{M}$, training cycles $\mathcal{C}$, diffusion steps $T$
\ENSURE Fine-tuned diffusion language model $\textit{DLM}$ for recommedation

\STATE Build the hybrid prompt dataset $\widetilde{\mathcal{D}}$ from $\mathcal{D}$, and record the positions of item-token blocks, textual description blocks, and target tokens for subsequent Stage 1 and Stage 2 masking

\FOR{$c = 1$ to $C$}
    \FOR{$n = 1$ to $\mathcal{N}$}
        \FOR{each batch $\mathcal{B} \subset \widetilde{\mathcal{D}}$}
            \STATE Sample $t \sim \mathrm{Uniform}(0,T)$
            \STATE Apply item-level curriculum masking to obtain masked prompts via Eq.~\ref{eq:diff_forward} and Eq. ~\ref{eq:item_level_mask_prob}
            \STATE Compute $\mathcal{L}_{\mathrm{diff}}$ via Eq.~\ref{eq:loss_diffusion_pre} and update $\textit{DLM}$
        \ENDFOR
    \ENDFOR

    \FOR{$m = 1$ to $\mathcal{M}$}
        \FOR{each batch $\mathcal{B} \subset \widetilde{\mathcal{D}}$}
            \STATE Sample $t \sim \mathrm{Uniform}(0,T)$ and set $p_{\mathrm{mask}}=t/T$
            \STATE Apply token-level masking to obtain corrupted prompts via Eq. ~\ref{eq:linear_diff_forward}
            \STATE Compute $\mathcal{L}_{\mathrm{diff}}$ via Eq.~\ref{eq:loss_diffusion_pre}
            \STATE Obtain the soft target representation $\hat{\mathbf{z}}$ from target-position logits via soft quantization in Eq.~\ref{eq:logit} and Eq.~\ref{eq:soft_rep}
            \STATE Construct $\langle \hat{\mathbf{z}}, \mathbf{z}_{\mathrm{pos}}, \mathbf{z}_{\mathrm{neg}} \rangle$ and compute $\mathcal{L}_{\mathrm{pref}}$ via Eq.~\ref{eq:loss_pref}
            \STATE Compute $\mathcal{L}_{\mathrm{S2}}$ via Eq.~\ref{eq:loss_stage2} and update $\textit{DLM}$
        \ENDFOR
    \ENDFOR
    \STATE Evaluate the model on the validation set
        \IF{early stopping criterion is satisfied}
        \RETURN $\textit{DLM}$        
        \ENDIF
\ENDFOR

\STATE \RETURN $\textit{DLM}$
\end{algorithmic}
\end{algorithm}

For masked positions, we retain the standard reconstruction objective $\mathcal{L}_{diff}$. However, recommendation requires more than token-level recovery, it demands that the model identify the true target item among competing alternatives. To this end, we introduce a preference-aware loss that contrasts the target item against hard negatives in the embedding space.
Since target prediction is defined over discrete codewords, hard selection blocks gradient flow and prevents the preference loss from backpropagating to the target-position logits. To address this, we adopt differentiable soft quantization to map the logits to a continuous item representation.
For each of the $K$ token positions, we extract the position-specific logits $\boldsymbol{\ell}^{(k)}$ and apply softmax to obtain a probability distribution $\mathbf{p}^{(k)} = \text{softmax}(\boldsymbol{\ell}^{(k)})$ over the corresponding sub-codebook. The expected codeword embedding is then computed as:
\begin{equation}
\tilde{\mathbf{c}}^{(k)} = \sum_{j=1}^{L} p_j^{(k)} \mathbf{c}_j^{(k)},
\label{eq:logit}
\end{equation}
where $p_j^{(k)}$ denotes the $j$-th component of $\mathbf{p}^{(k)}$, representing the probability of selecting the $j$-th codeword in the $k$-th sub-codebook. The $K$ expected embeddings are then concatenated and passed through the decoder to reconstruct the expected item representation: 
\begin{equation}
\hat{\mathbf{z}} = f_{\text{dec}}\left([\tilde{\mathbf{c}}^{(1)} , \tilde{\mathbf{c}}^{(2)} \ldots \tilde{\mathbf{c}}^{(K)}]\right).
\label{eq:soft_rep}
\end{equation}
This differentiable decoding process allows the preference loss to back-propagate to the target-position logits while remaining compatible with discrete codeword prediction.
To provide recommendation-oriented supervision,, we construct preference triplets $\langle \hat{\mathbf{z}}, \mathbf{z}_{\text{pos}}, \mathbf{z}_{\text{neg}} \rangle$ in the embedding space. 
The positive anchor $\mathbf{z}_{\text{pos}}$ is the ground-truth embedding of the target item.
For the negative sample, we perform hard negative mining based on cosine similarity: we compute the similarity between $\hat{\mathbf{z}}$ and all items in the embedding bank $\mathbf{Q}$, exclude the target item and historical interacted items, and randomly sample one from the top-$5$ most similar most similar remaining candidates as as $\mathbf{z}_{\text{neg}}$. The preference loss is defined as a pairwise ranking objective:
\begin{equation}
\mathcal{L}_{\text{pref}} = -\log \sigma(\cos(\hat{\mathbf{z}}, \mathbf{z}_{\text{pos}}) - \cos(\hat{\mathbf{z}}, \mathbf{z}_{\text{neg}})),
\label{eq:loss_pref}
\end{equation}
where $\cos(\cdot,\cdot)$ denotes cosine similarity, and $\sigma(\cdot)$ is the sigmoid function. This objective encourages the decoded representation to stay closer to the true target item than to hard negatives, thereby improving the model’s discriminative ability for recommendation.
The overall Stage-2 objective combines the diffusion reconstruction loss with the proposed preference-aware loss:
\begin{equation}
\mathcal{L}_{\mathrm{S2}}
=
\mathcal{L}_{\mathrm{diff}}
+
w_{\mathrm{pref}}
\mathcal{L}_{\mathrm{pref}},
\label{eq:loss_stage2}
\end{equation}
where $w_{\mathrm{pref}}$ controls the contribution of the preference-aware term.


The overall training framework alternates between Stage 1 and Stage 2 over multiple cycles, where Stage 1 runs for $\mathcal{N}$ epochs with progressive item-level masking and Stage 2 runs for $\mathcal{M}$ epochs with preference-aware token-level denoising; detailed procedures are summarized in Algorithm ~\ref{alg:two_stage_training}.

\subsection{Voting-based Iterative Refinement for Recommendation}

Standard discrete diffusion sampling typically relies on re-masking and regenerating from scratch, which may discard informative intermediate preference signals and destabilize recommendation outputs. To address this, we propose a voting-based iterative refinement strategy tailored to discrete item tokens. By aggregating token-level probability mass across refinement steps, the model selects candidates that remain consistently supported throughout generation. This mechanism preserves useful intermediate information and enables smoother transitions from noisy states to high-fidelity tokens, improving both stability and reliability of inference.

\subsubsection{Prediction-guided Iterative Refinement}
To enable gradual refinement of recommendation outputs, we introduce a prediction-guided iterative refinement strategy, building upon recent advances in diffusion decoding~\cite{pynadath2025candi,zheng2025continuously}. In contrast to conventional approaches that reset target tokens to masked states at each denoising step, DLMRec preserves information from prior predictions within the continuous embedding space. Concretely, the predicted target tokens are projected back to their corresponding codeword embeddings (i.e., the learnable input representation within the DLM), which are then aggregated with mask embeddings and user-conditioned collaborative representations to construct the input for the subsequent refinement step. This iterative design facilitates the accumulation of recommendation-relevant evidence throughout the generation trajectory, resulting in more accurate and preference-aligned predictions.

At each refinement step, the logits at target positions are projected into the DLM input embedding space for continuous refinement. We consider two modes: hard and soft. The hard mode uses the input embedding of the argmax token, while the soft mode computes a probability-weighted average over the target-position token distribution.
Formally, for the $k$-th target position with logit vector $\boldsymbol{\ell}^{(k),i} \in \mathbb{R}^{L}$ at refinement step $i$, let
\begin{equation}
\hat{\mathbf{e}}^{(k),i} =
\begin{cases}
\mathbf{e}^{(k)}_{j^*},\quad j^* = \arg\max\limits_{j=1,\ldots,L} \ell^{(k),i}_j, \quad & \text{(hard)}, \\[6pt]
\sum\limits_{l=1}^{L} \mathbf{p}^{(k),i}_l \, \mathbf{e}^{(k)}_l,  \mathbf{p}^{(k),i}_l = \text{softmax}(\boldsymbol{\ell}^{(k),i})_l, \quad & \text{(soft)},
\end{cases}
\label{eq:refinement_representation}
\end{equation}
where $\mathbf{e}^{(k)}_l$ denotes the input embedding of the $l$-th candidate token in the DLM embedding layer for the $k$-th target position.
To guide subsequent denoising, we first construct a user-conditioned prior embedding for each target hop:
\begin{equation}
\mathbf{b}^{(k)}=\mathbf{e}_{\mathrm{mask}}+\lambda \mathbf{e}_{u}^{(k)},
\label{eq:user_conditioned_prior}
\end{equation}
where $\mathbf{e}_{\mathrm{mask}}$ denotes the DLM input embedding of the mask token and $\mathbf{e}_{u}^{(k)}$ is the DLM input embedding of the user's $k$-th token. The mask embedding serves as a neutral placeholder consistent with the DLM decoding process, while the user embedding injects collaborative preference information. Their combination therefore provides a user-conditioned prior for refining uncertain target positions.
The estimated embedding at target hop $k$, $\hat{\mathbf{e}}^{(k),i}$, is then interpolated with this prior to form the input of the next refinement step:
\begin{equation}
\begin{aligned}
\mathbf{h}^{(k),i+1}
&=(1-\alpha^{(k),i})\hat{\mathbf{e}}^{(k),i}+\alpha^{(k),i}\mathbf{b}^{(k)}, \\
\alpha^{(k),i}
&=\sigma\!\left(\tau-c^{(k),i}\right),
\end{aligned}
\label{eq:update}
\end{equation}
where $\alpha^{(k),i}$ is the confidence-guided mixing coefficient, $c^{(k),i}$ denotes the maximum softmax probability over the codebook logits at hop $k$ during refinement step $i$, and $\tau$ is a confidence threshold. 
Thus, confident predictions are preserved through their estimated embeddings, while uncertain positions are pulled toward the user-conditioned prior. The resulting embedding $\mathbf{h}^{(k),i+1}$ is then fed into the DLM for the next refinement iteration.
To ensure training–inference consistency, we also replace the masked target item tokens in Stage 2 with the same user-conditioned prior embedding in Eq. ~\ref{eq:loss_diffusion_pre}, rather than using the plain mask token alone. This allows the model to adapt to the user-guided refinement pattern.

\subsubsection{Stability-aware Voting for Recommendation}
Although refinement preserves intermediate information, it does not ensure uniform convergence across all target positions. In practice, some positions stabilize early with consistent predictions, while others remain uncertain and oscillate across consecutive iterations. Treating all positions uniformly risks overwriting reliable predictions and amplifying drift. 
To address this, we introduce a stability-aware voting strategy that adaptively handles positions based on their convergence behavior. Stable positions are committed early to their argmax tokens once convergence is detected, while unstable ones continue soft refinement under user-conditioned guidance. At the end of inference, voting aggregates evidence from the refinement trajectory to produce robust final recommendations by leveraging consensus across steps.

Specifically, after each refinement step, we compare the current prediction at hop 
$k$ with that from the previous step, and define its stability as:
\begin{equation}
\mathrm{stable}_k^{(i)}
=
\mathbf{1}\!\left[
\arg\max \mathbf{p}^{(k),i} = \arg\max \mathbf{p}^{(k),i-1}
\;\land\;
\max \mathbf{p}^{(k),i} > \gamma
\right],
\end{equation}
where $\mathbf{p}^{(k),i}$ denotes the token distribution at hop $k$ in refinement step $i$, and $\gamma$ is a confidence threshold. A target position is regarded as stable if its predicted token remains unchanged and the corresponding confidence exceeds $\gamma$; otherwise, it is treated as unstable. For stable positions, the predicted token is mapped to its DLM input embedding and directly used as a hard replacement in the next refinement step. For unstable positions, we continue to apply the prediction-guided interpolation update described in Eq. ~\ref{eq:update}.

At the final step, stable positions are directly decoded using their argmax tokens. For unstable positions, we apply voting to aggregate evidence across refinement steps: log-probabilities from the current and previous steps are summed, and the token with the highest cumulative score is selected. This favors the most consistently supported candidate across the trajectory, improving prediction stability. The finalized token sequence is then decoded into a continuous representation for nearest-neighbor retrieval over the item bank to produce the top-$\mathcal{K}$ recommendation list as demonstrated in Algorithm ~\ref{alg:inference_sampling}.

\begin{algorithm}[t]
\caption{Voting-based Iterative Refinement for Recommendation}
\label{alg:inference_sampling}
\begin{algorithmic}[1]
\REQUIRE Input embeddings $\mathbf{x}_{S}$ with target item positions masked, diffusion language model $\textit{DLM}$, CAST decoder $f_{\mathrm{dec}}$, item representations $\mathbf{Q}$, refinement steps $S$, user guidance weight $\lambda$, confidence threshold $\gamma$, mixing threshold $\tau$
\ENSURE Top-$N$ recommendation list

\STATE Perform an initial forward pass on $\mathbf{x}_{S}$ to obtain target-position logits $\{\boldsymbol{\ell}^{(k),0}\}_{k=1}^{K}$
\STATE Extract initial target distributions $\{\mathbf{p}^{k,(0)}\}_{k=1}^{K}$, argmax indices, confidence scores, and log-probabilities
\IF{$S=1$}
    \STATE Set the final CAST token at hop $k$ as $z_k^\ast = \arg\max \mathbf{p}^{k,(0)}$ for all $k$
\ELSE
    \FOR{$i=1$ to $S-1$}
        \FOR{$k=1$ to $K$}
            \STATE Compute confidence score $c^{(k),i} = \max \mathbf{p}^{(k),i}$

            \IF{$\arg\max \mathbf{p}^{(k),i} = \arg\max \mathbf{p}^{(k),i-1}$ \AND $c^{(k),i} > \gamma$}
                \STATE Mark hop $k$ as stable and replace its representation with the corresponding DLM input embedding of the argmax token
            \ELSE
                \STATE Mark hop $k$ as unstable and construct next-step representation by combining prediction embedding (Eq.~\ref{eq:refinement_representation}) and user-conditioned base embedding (Eq.~\ref{eq:user_conditioned_prior}) via Eq.~\ref{eq:update}
            \ENDIF
        \ENDFOR
        \STATE Update the target-region embeddings with the refined representations and feed them into $\textit{DLM}$ via \texttt{inputs\_embeds}
        \STATE Obtain new target logits $\{\boldsymbol{\ell}^{(k),i+1}\}_{k=1}^{K}$ and update $\{\mathbf{p}^{(k),i+1}\}_{k=1}^{K}$, log-probabilities, and prediction embeddings
    \ENDFOR

    \FOR{$k=1$ to $K$}
        \IF{hop $k$ is stable at the final step}
            \STATE Set $z_k^\ast = \arg\max \mathbf{p}_k^{(S-1)}$
        \ELSE
            \STATE Perform voting over all refinement steps: $ z_k^\ast = \arg\max \sum_{i=0}^{S-1} \log \mathbf{p}_k^{(i)}$
        \ENDIF
    \ENDFOR
\ENDIF

\STATE Form the finalized CAST token sequence $\mathbf{z}^\ast = [z_1^\ast,\ldots,z_K^\ast]$
\STATE Decode $\mathbf{z}^\ast$ into the target item representation $\hat{\mathbf{z}} = f_{\mathrm{dec}}(\mathbf{z}^\ast)$
\STATE Retrieve the top-$N$ most similar items from the item bank $\mathbf{Q}$ using $\hat{\mathbf{z}}$
\RETURN Top-$N$ recommendation list
\end{algorithmic}
\end{algorithm}

\subsection{Complexity Analysis of DLMRec}

In this section, we analyzes the time complexity of the proposed DLMRec framework.

\subsubsection{Complexity of Training}

Our model is trained using a masked diffusion objective with bidirectional attention over the full sequence. Let $L_s$ denote the sequence length, $H$ the number of Transformer layers, and $d$ the hidden dimension. For each training sample, the computational cost is dominated by the Transformer forward computation, which incurs a complexity of $\mathcal{O}(H \cdot L_s^2 \cdot d)$.
In addition, our curriculum-based masking strategy gradually transitions from item-level to token-level masking. At each training step, the masking operation introduces a linear overhead of $\mathcal{O}(L_s)$ for constructing corrupted inputs. Therefore, the overall training complexity per sample is $\mathcal{O}(H \cdot L_s^2 \cdot d + L_s) \approx \mathcal{O}(H \cdot L_s^2 \cdot d)$, where the masking cost is negligible compared to the Transformer computation.
Compared to autoregressive training, which optimizes a factorized likelihood and relies on causal masking, our method removes the autoregressive constraint and enables full parallel computation over all positions. As a result, both approaches have comparable asymptotic training complexity in terms of FLOPs, while our method benefits from bidirectional context modeling without incurring additional computational cost.

\subsubsection{Complexity of Inference}

At inference time, DLMRec performs $S$ iterative refinement steps over the full input sequence. In each step, a forward pass with bidirectional attention is executed, incurring a complexity of $\mathcal{O}(S \cdot H \cdot L_s^2 \cdot d)$ for the Transformer computation. 
Unlike standard decoding strategies that regenerate all positions at each step, our stability-aware mechanism operates selectively: stable token positions are committed early and fixed without further refinement, while only unstable positions ($K$ targets, where $K \ll L_s$) undergo continued updates via embedding interpolation and confidence-based gating. The overhead from stability checking, embedding updates, and adaptive interpolation restricted to the target region is $\mathcal{O}(S \cdot K \cdot d)$, which is negligible compared to the Transformer forward pass. The voting mechanism aggregates log-probabilities across refinement steps exclusively for unstable positions, introducing an additional cost of $\mathcal{O}(K \cdot S)$, which is also negligible since both $K$ and $S$ are small constants. 
The overall inference complexity is dominated by Transformer forward computation: $\mathcal{O}(S \cdot H \cdot L_s^2 \cdot d)$.


%% file: sections/Experiment.tex
\section{Experiment}
\label{Experiment}

In this section, we present comprehensive experiments to evaluate the proposed DLMRec framework.

\subsection{Experiment Settings}
\subsubsection{Datasets}

We conduct experiments on three benchmark datasets, including LastFM ~\footnote{LastFM: \url{https://grouplens.org/datasets/hetrec-2011/}}, MovieLens-1M ~\footnote{MovieLens-1M: \url{https://grouplens.org/datasets/movielens/1m/}}, and Amazon-Beauty~\footnote{Amazon-Beauty: \url{https://nijianmo.github.io/amazon/}}, to evaluate the effectiveness of our proposed DLMRec framework across different recommendation scenarios.
LastFM, obtained from the HetRec 2011 repository, records user listening histories of music artists. MovieLens-1M consists of movie ratings provided by MovieLens users. Amazon-Beauty, sourced from the Amazon Review Data repository, contains user reviews and ratings for beauty products.
Following common practice, we filter out users and items with fewer than three interactions. As textual side information, we use item titles for MovieLens-1M and Amazon-Beauty, and artist names for LastFM. User and item IDs are retained as collaborative identifiers across all datasets. We adopt the Split-by-Timepoint protocol~\cite{ji2023critical}, sorting interactions chronologically and using those before a cutoff timestamp for training and the remaining ones for testing. Table~\ref{tab:dataset_stats} summarizes the resulting dataset statistics.

\input{tables/dataset_stats}

\subsubsection{Evaluation Metrics}
We evaluate top-$\mathcal{K}$ recommendation performance using Recall@$\mathcal{K}$ and Normalized Discounted Cumulative Gain (NDCG@$\mathcal{K}$) with $\mathcal{K}\in \{10,20 \}$ under the full-ranking setting. All reported results are averaged over five independent runs.

\subsubsection{Baselines}
We compare the proposed DLMRec with a broad range of competitive baselines categorized into four groups: GNN-based collaborative filtering (LightGCN~\cite{he2020lightgcn}, SGL~\cite{wu2021self}, LTGNN~\cite{zhang2024linear}), sequential recommendation (SASRec~\cite{kang2018self}, BERT4Rec~\cite{sun2019bert4rec}), diffusion-based generative recommendation (DiffRec~\cite{wang2023diffusion}, CDRec~\cite{liu2026continuous}, LLaDaRec~\cite{shi2025llada}), and LLM-empowered recommenders (TIGER~\cite{rajput2023recommender}, LLaRa~\cite{liao2024llara}, CoLLM~\cite{zhang2025collm}, TokenRec~\cite{qu2025tokenrec}). We briefly describe each baseline below.

\begin{itemize}[leftmargin=*]
\item LightGCN~\cite{he2020lightgcn}: LightGCN refines graph convolutional networks by eliminating feature transformation and nonlinearities, relying solely on the neighborhood aggregation process for collaborative filtering.
\item SGL~\cite{wu2021self}: SGL generates multiple views of the interaction graph and employs contrastive learning to enforce consistency across them, thereby enhancing the supervisory signal.
\item LTGNN~\cite{zhang2024linear}: LTGNN achieves linear-time GNN-based recommendation via implicit graph modeling and variance-reduced sampling.
\item SASRec~\cite{kang2018self}: SASRec applies self-attention to sequential recommendation, adaptively weighting previous items to predict the next item.
\item BERT4Rec~\cite{sun2019bert4rec}: BERT4Rec employs deep bidirectional self-attention to sequential recommendation, enabling each item to fuse both left and right context.
\item DiffRec~\cite{wang2023diffusion}: DiffRec employs a diffusion-based denoising approach that perturbs interaction vectors and learns to reconstruct them for recommendation.
\item CDRec~\cite{liu2026continuous}: CDRec introduces discrete-space diffusion to recommendation via popularity-aware masking in continuous time, enabling efficient single-step generation with consistency parameterization.
\item LLaDaRec~\cite{shi2025llada}: LLaDA-Rec adapts discrete diffusion to generative recommendation with parallel semantic IDs and adaptive beam search.
\item TIGER~\cite{rajput2023recommender}: TIGER compresses textual information into compact semantic IDs via residual vector quantization, and trains a Transformer to model semantic ID sequences for sequential recommendation.
\item LLaRa~\cite{liao2024llara}: LLaRa leverages a projection module to bridge item embeddings and textual item descriptions, and further applies curriculum learning during fine-tuning.
\item CoLLM~\cite{zhang2025collm}: CoLLM adopts GNNs to encode users and items into continuous embeddings, which are then fed into LLMs for recommendation.
\item TokenRec~\cite{qu2025tokenrec}: TokenRec introduces a quantized tokenizer that discretizes users and items with their collaborative semantics into tokens compatible with LLMs.
\end{itemize}

\subsubsection{Hyper-parameter Setting}
The DLRMRec framework is built on PyTorch and the Hugging Face platform, with LLaDA-8B-Base~\footnote{\url{https://huggingface.co/GSAI-ML/LLaDA-8B-Base}}~\cite{nie2025large} serving as the diffusion language model backbone. 
We apply a linear warmup-decay learning rate schedule (\texttt{get\_linear\_schedule\_with\_warmup}) and employ LoRA~\cite{hu2022lora} for parameter-efficient fine-tuning.
The codebook size $K$ is tuned from $\{1,2,3,4\}$, and the number of codewords $L$ is selected from $\{64,128,256,512\}$. The top-$\mathcal{S}$ candidates for stochastic assignment are set in $\{2,3\}$. The lower and upper temperature bounds $\tau_{\min}$ and $\tau_{\max}$ are chosen from $\{0.05,0.5\}$ and $\{0.1,1.0\}$, respectively, and the growth rate $\alpha$ is set to $1.1$ in Eq.~\ref{eq:transition_distribution} for all experiments.
During sampling, the number of refinement iterations is selected from $\{1,2,3,4,5,10\}$ while the confidence threshold $\tau$ is tuned from $\{0.1,0.3,0.5\}$.
To ensure a fair comparison, we employ Llama-3-8B-Base as the backbone autoregressive LLM across all baselines. For LLM-based approaches, we adopt the same prompting templates and fine-tuning configurations as reported in their respective papers, and carefully calibrate hyperparameters to account for differences in model scale.

\subsection{Overall Performance}

\input{tables/perf}

We conduct a thorough evaluation of the proposed DLMRec framework in comparison with various state-of-the-art baselines. 
Table ~\ref{OverallPerfromance} presents the Recall and NDCG performance across three benchmark datasets, LastFM, MovieLens-1M, and Amazon-Beauty. The main findings are as follows:
\begin{itemize}[leftmargin=*]
\item The proposed DLMRec demonstrates superior effectiveness, consistently outperforming all representative baselines across datasets. Compared with the strongest LLM-based recommender system (LLaRA), DLMRec achieves average improvements of $6.75\%$ and $5.19\%$ in Recall and NDCG, respectively. When compared with the discrete diffusion baseline, the gains further increase to $11.83\%$ and $6.37\%$. These improvements validate the effectiveness of DLMRec in bridging collaborative semantics and language modeling, as well as the benefits of iterative diffusion-based generation for recommendation.

\item Overall, LLM-based recommenders outperform both traditional collaborative filtering and diffusion-based methods across datasets, highlighting the advantage of leveraging rich semantic information and pre-trained language models for preference modeling. By incorporating textual item descriptions and user interaction context, these methods jointly model semantic and behavioral signals, enabling a more comprehensive understanding of user preferences

\item Diffusion-based methods consistently outperform traditional discriminative deep learning models across all datasets, highlighting the effectiveness of generative modeling for recommendation. More importantly, diffusion-based approaches are also competitive with strong LLM-based recommenders. This indicates that iterative denoising provides a powerful and flexible generative paradigm for modeling complex user--item interaction patterns, underscoring the potential of diffusion models for recommendation tasks.

\item GNN-based methods typically outperform the sequential baselines across all three datasets, demonstrating their effectiveness in capturing both direct and higher-order collaborative signals in the user-item interaction graph. This motivates the use of a GNN encoder to construct user and item representations prior to tokenization, ensuring that discrete tokens are grounded in collaborative semantics derived from the interaction graph structure.
\end{itemize}

\subsection{Ablation Study}

\begin{figure}[tp]
    \centering
    \begin{subfigure}[b]{0.32\textwidth}
        \includegraphics[width=\textwidth]{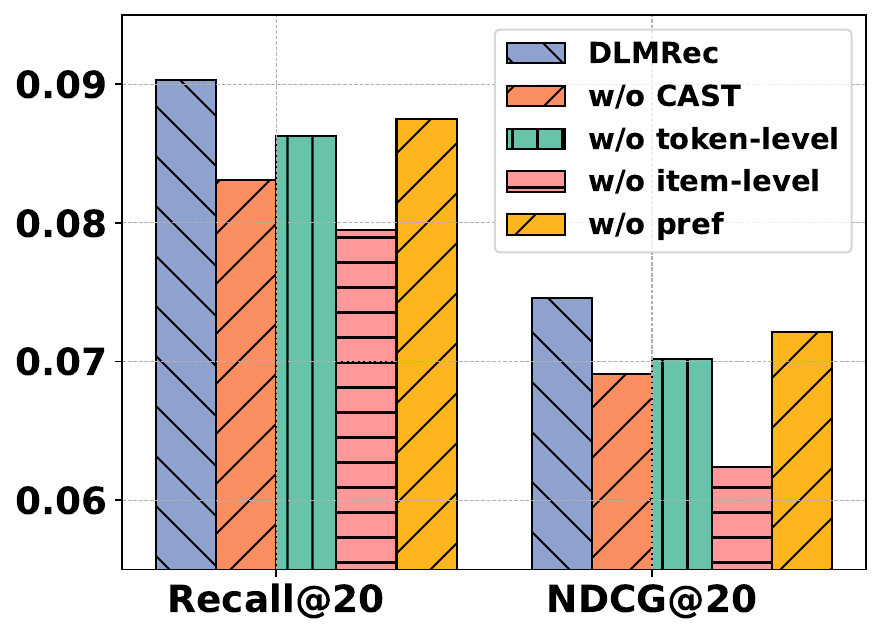}
        \caption{LastFM}
    \end{subfigure}
    \begin{subfigure}[b]{0.32\textwidth}
        \includegraphics[width=\textwidth]{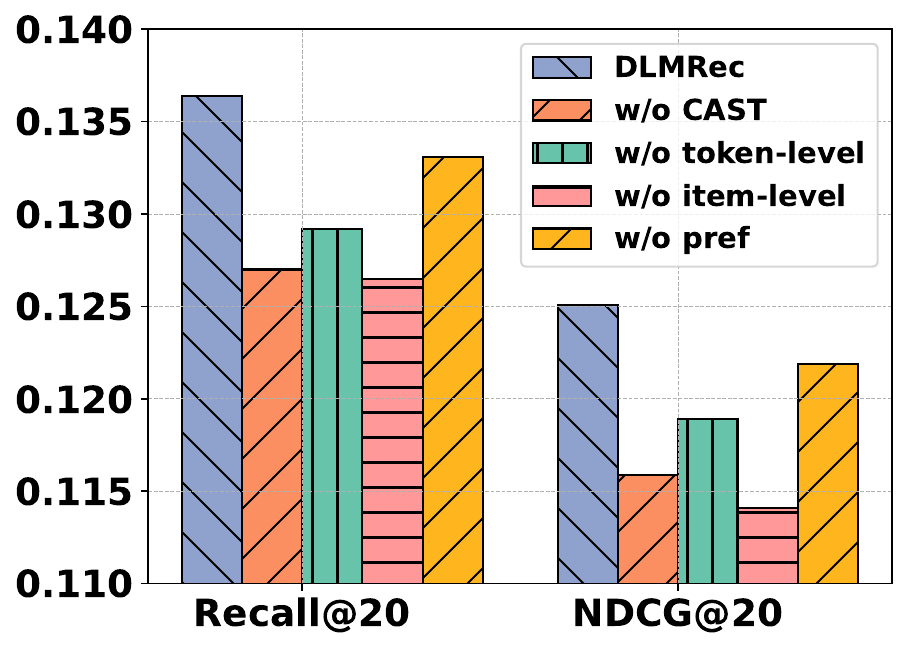}
        \caption{MovieLens-1M}
    \end{subfigure}
    \begin{subfigure}[b]{0.32\textwidth}
        \includegraphics[width=\textwidth]{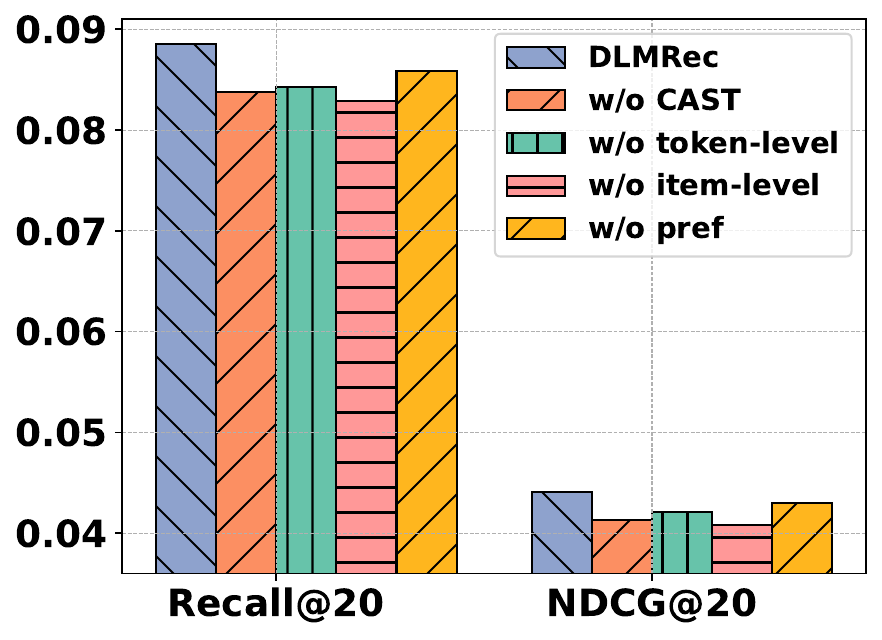}
        \caption{Amazon-Beauty}
    \end{subfigure}
    
    \begin{subfigure}[b]{0.32\textwidth}
        \includegraphics[width=\textwidth]{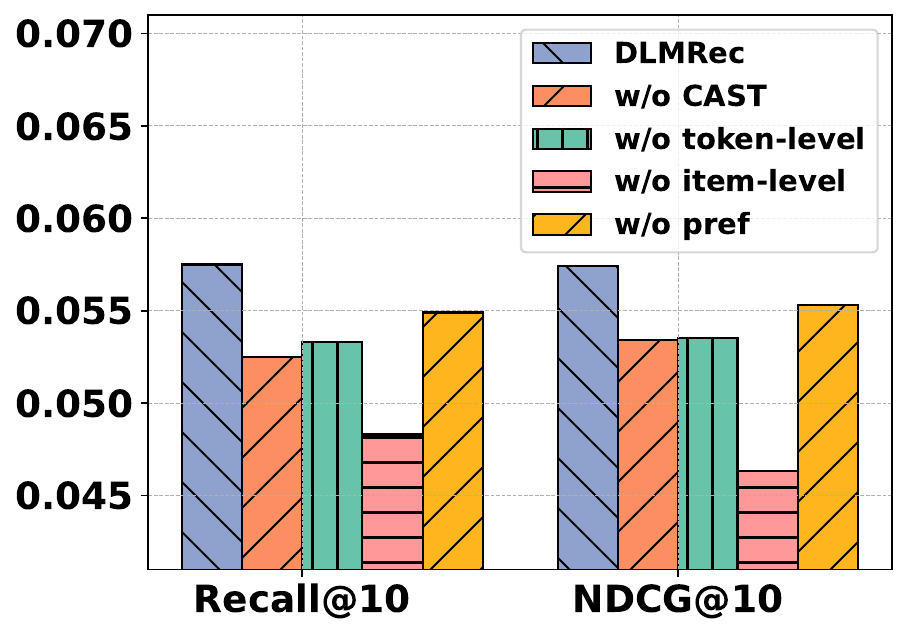}
        \caption{LastFM}
    \end{subfigure}
    \begin{subfigure}[b]{0.32\textwidth}
        \includegraphics[width=\textwidth]{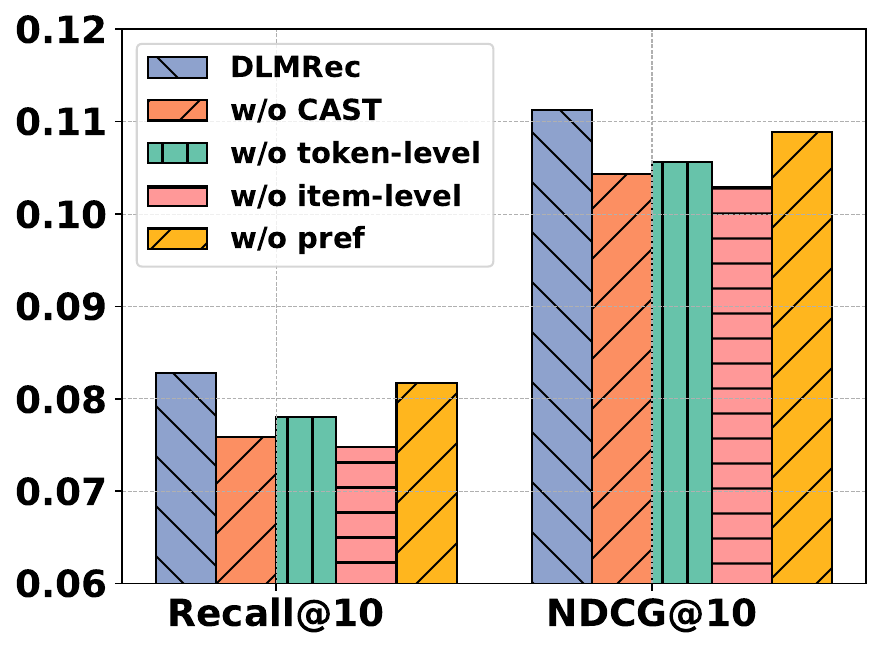}
        \caption{MovieLens-1M}
    \end{subfigure}
    \begin{subfigure}[b]{0.32\textwidth}
        \includegraphics[width=\textwidth]{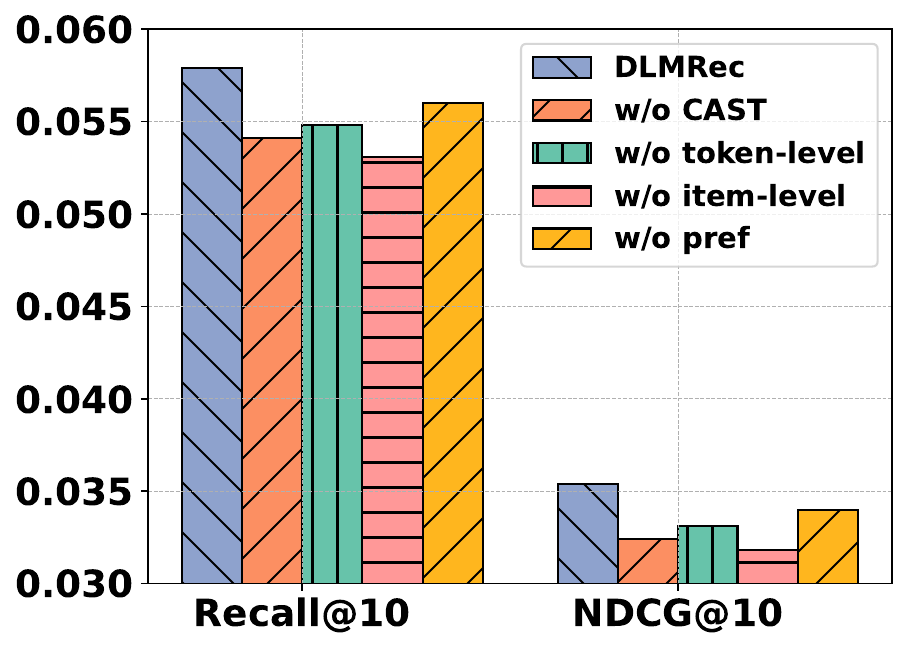}
        \caption{Amazon-Beauty}
    \end{subfigure}
    \caption{Ablation study of DLMRec and its variants on three
    datasets }
    \label{fig:ablation}
    \Description{ablation}
\end{figure}

To systematically assess the impact of each key component, we construct four variants of DLMRec by ablating specific modules.

\begin{itemize}[leftmargin=*]
\item \textit{w/o CAST}: This variant ablates the collaborative-aware stochastic tokenizer and adopts standard VQ-VAE~\cite{van2017neural} for quantization.
\item \textit{w/o token-level}: This variant disables token-level masking in the preference-aware denoising stage, relying on item-level masking for training.
\item \textit{w/o item-level}: \textit{w/o item-level}: Removes item-level masking from the token-semantic alignment stage and adopts random token masking during training.
\item \textit{w/o pref}: This variant eliminates the preference loss in Eq. ~\ref{eq:loss_pref} and models user preferences solely using the standard masked reconstruction objective.
\end{itemize}

The ablation study results are presented in Fig.~\ref{fig:ablation}. CAST consistently outperforms the \textit{w/o CAST} variant across all three datasets, validating the effectiveness of stochastic token assignment in preserving collaborative semantics and uncertainty for more expressive discrete representations.
Removing either token-level or item-level masking leads to noticeable performance degradation, with the most severe drop observed when item-level masking is ablated. 
This highlights the necessity of item-level masking for coarse-to-fine preference alignment, validating the curriculum learning strategy. It also confirms its role or token-level masking in semantic-level guidance for preference-aware denoising.
The preference loss consistently improves performance, confirming its effectiveness in providing ranking-oriented supervision by contrasting predictions against ground-truth targets and hard negatives to enhance recommendation discriminability.

\subsection{Parameter Sensitivity Study}
In this section, we evaluate the impact of the key parameters in DLMRec.

\subsubsection{Effect of Codebook Settings}

\begin{figure}[tp]
    \centering
    \begin{subfigure}[b]{0.245\textwidth}
        \includegraphics[width=\textwidth]{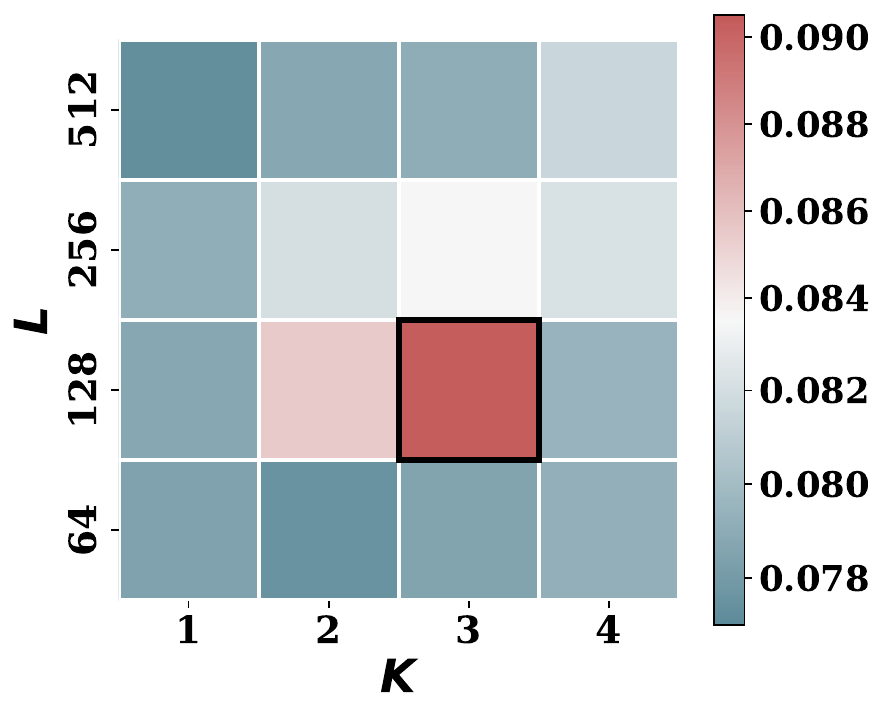}
        \caption{LastFM-Recall@20}
    \end{subfigure}
    \begin{subfigure}[b]{0.245\textwidth}
        \includegraphics[width=\textwidth]{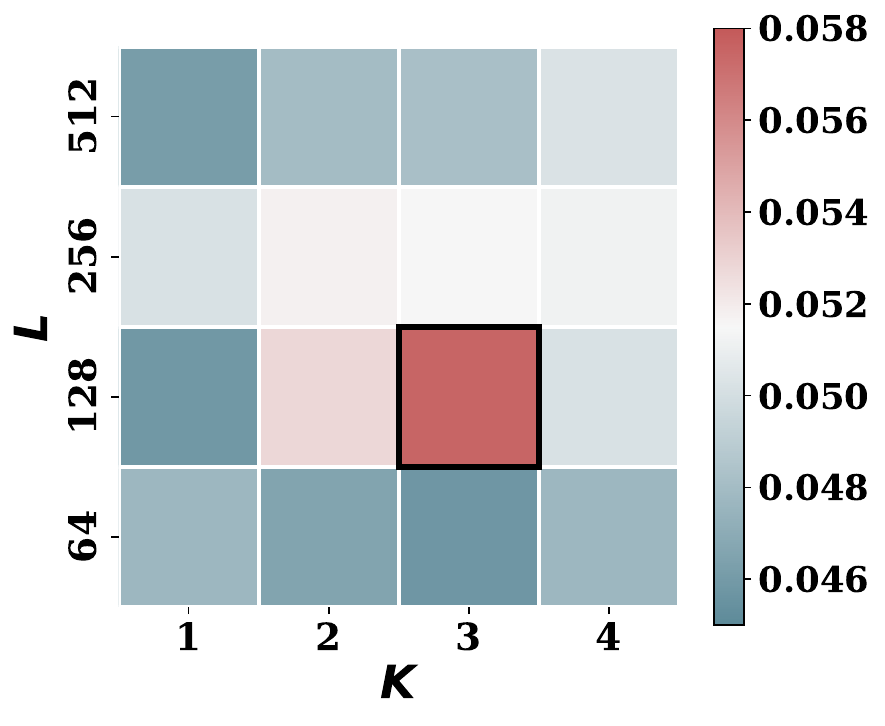}
        \caption{LastFM-Recall@10}
    \end{subfigure}
    \begin{subfigure}[b]{0.245\textwidth}
        \includegraphics[width=\textwidth]{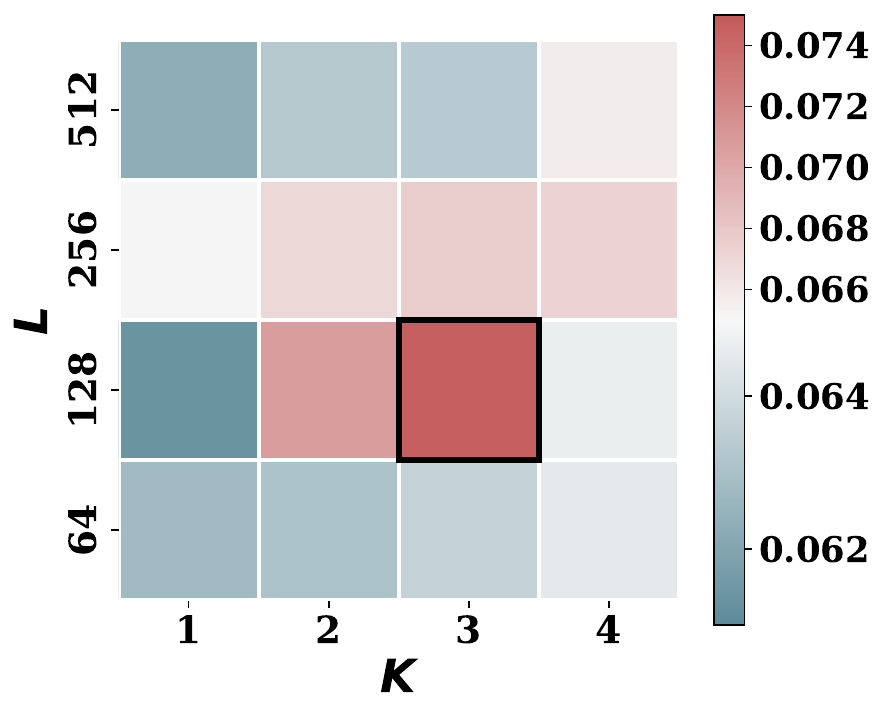}
        \caption{LastFM-NDCG@20}
    \end{subfigure}
    \begin{subfigure}[b]{0.245\textwidth}
        \includegraphics[width=\textwidth]{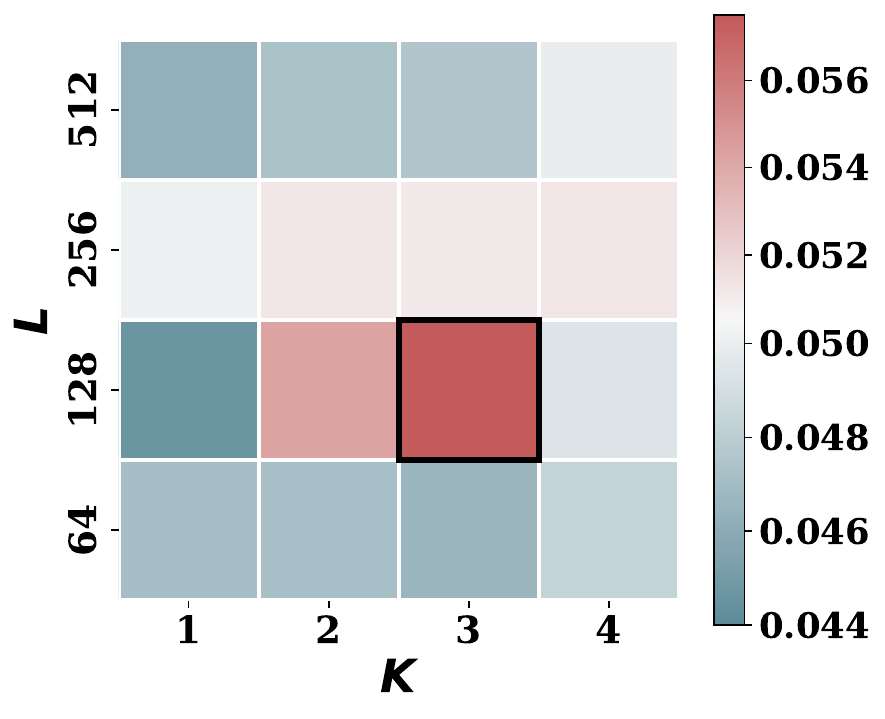}
        \caption{LastFM-NDCG@10}
    \end{subfigure}

    \begin{subfigure}[b]{0.245\textwidth}
        \includegraphics[width=\textwidth]{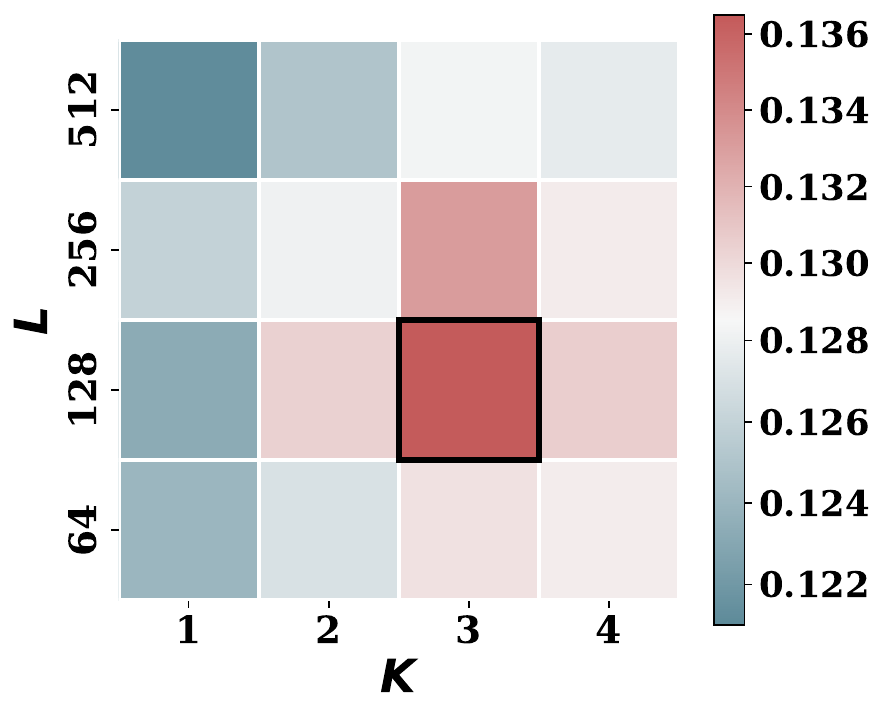}
        \caption{MovieLens-1M-Recall@20}
    \end{subfigure}
    \begin{subfigure}[b]{0.245\textwidth}
        \includegraphics[width=\textwidth]{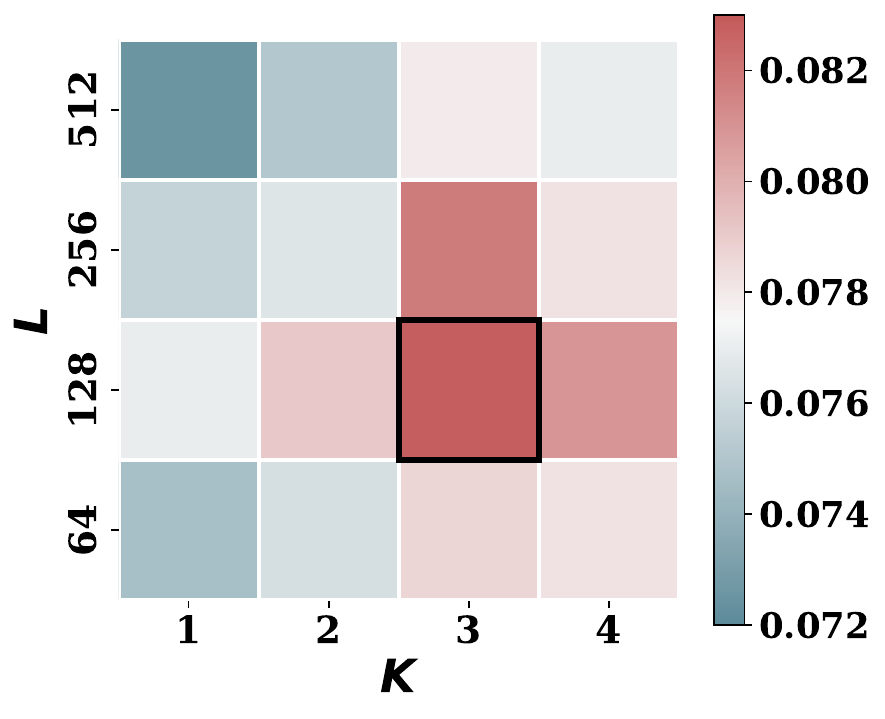}
        \caption{MovieLens-1M-Recall@10}
    \end{subfigure}
    \begin{subfigure}[b]{0.245\textwidth}
        \includegraphics[width=\textwidth]{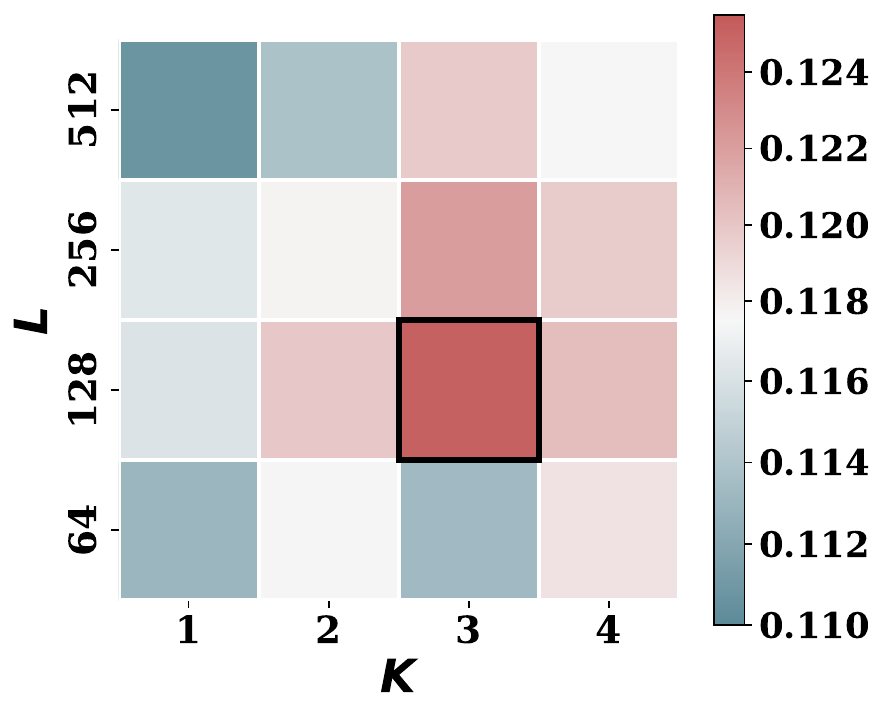}
        \caption{MovieLens-1M-NDCG@20}
    \end{subfigure}
    \begin{subfigure}[b]{0.245\textwidth}
        \includegraphics[width=\textwidth]{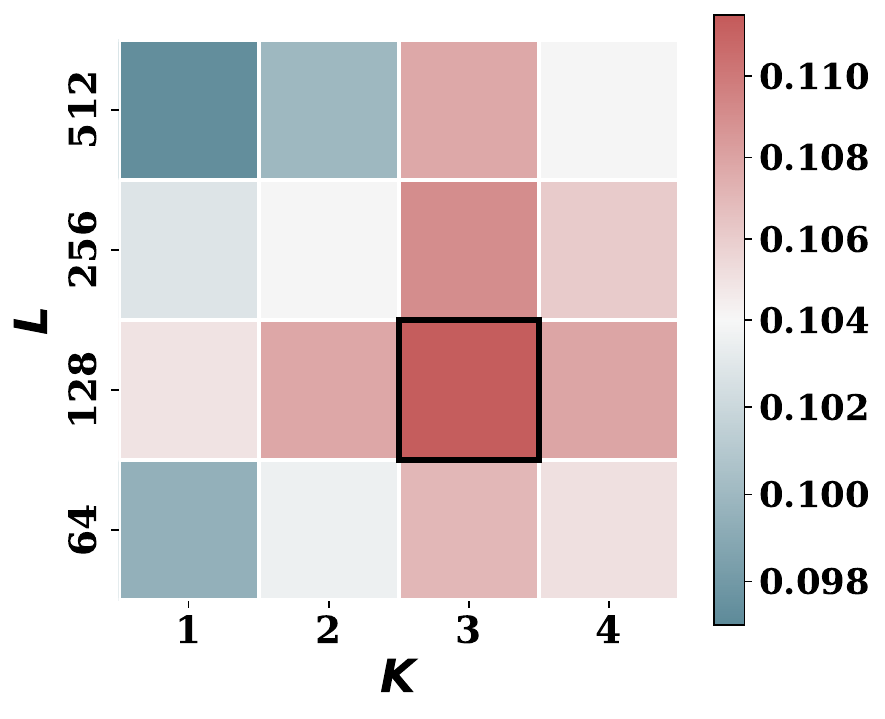}
        \caption{MovieLens-1M-NDCG@10}
    \end{subfigure}

        \begin{subfigure}[b]{0.245\textwidth}
        \includegraphics[width=\textwidth]{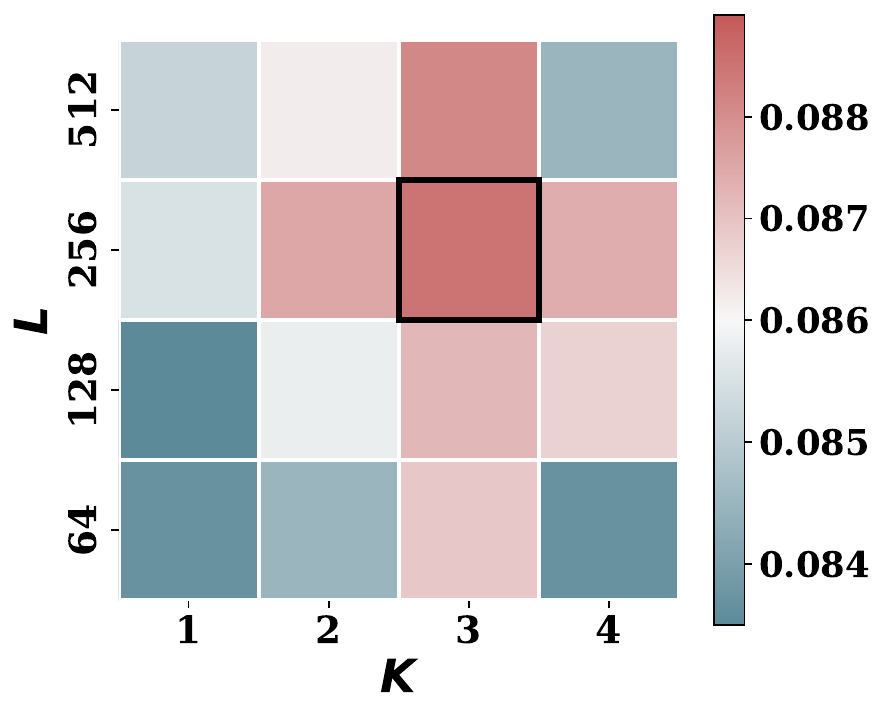}
        \caption{Amazon-Beauty-Recall@20}
    \end{subfigure}
    \begin{subfigure}[b]{0.245\textwidth}
        \includegraphics[width=\textwidth]{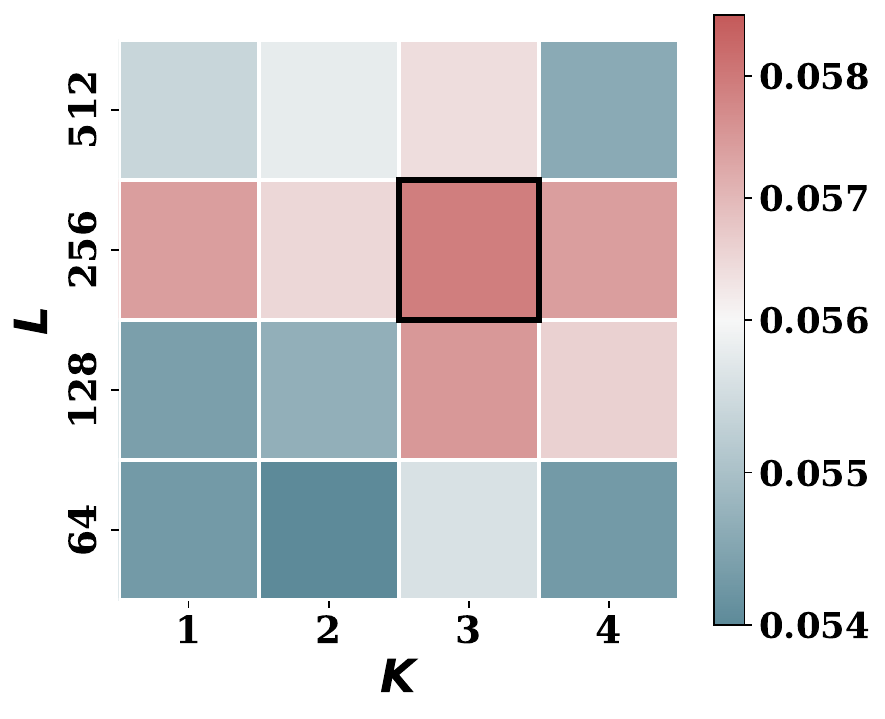}
        \caption{Amazon-Beauty-Recall@10}
    \end{subfigure}
    \begin{subfigure}[b]{0.245\textwidth}
        \includegraphics[width=\textwidth]{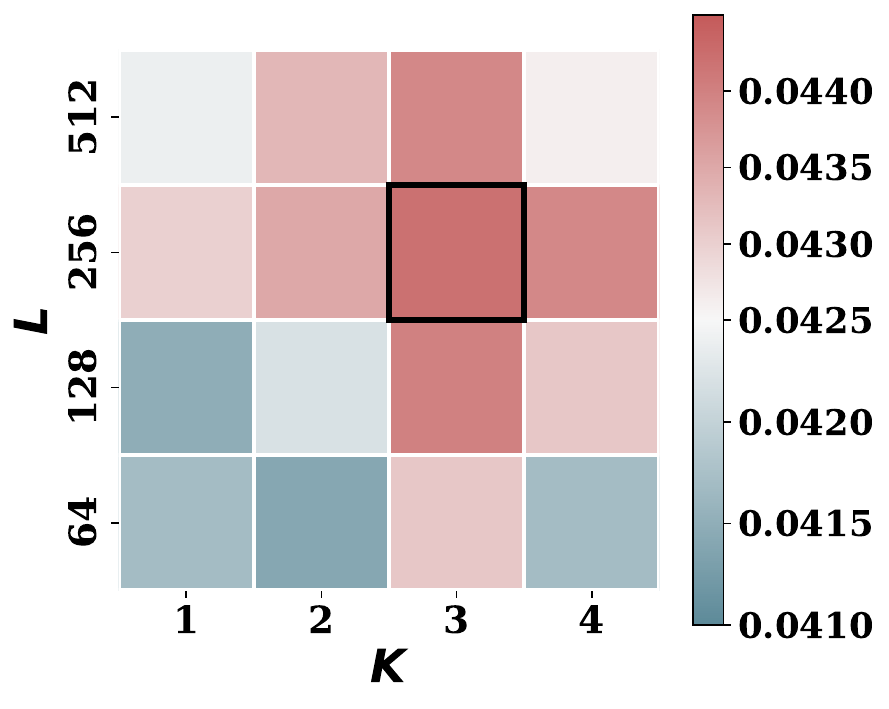}
        \caption{Amazon-Beauty-NDCG@20}
    \end{subfigure}
    \begin{subfigure}[b]{0.245\textwidth}
        \includegraphics[width=\textwidth]{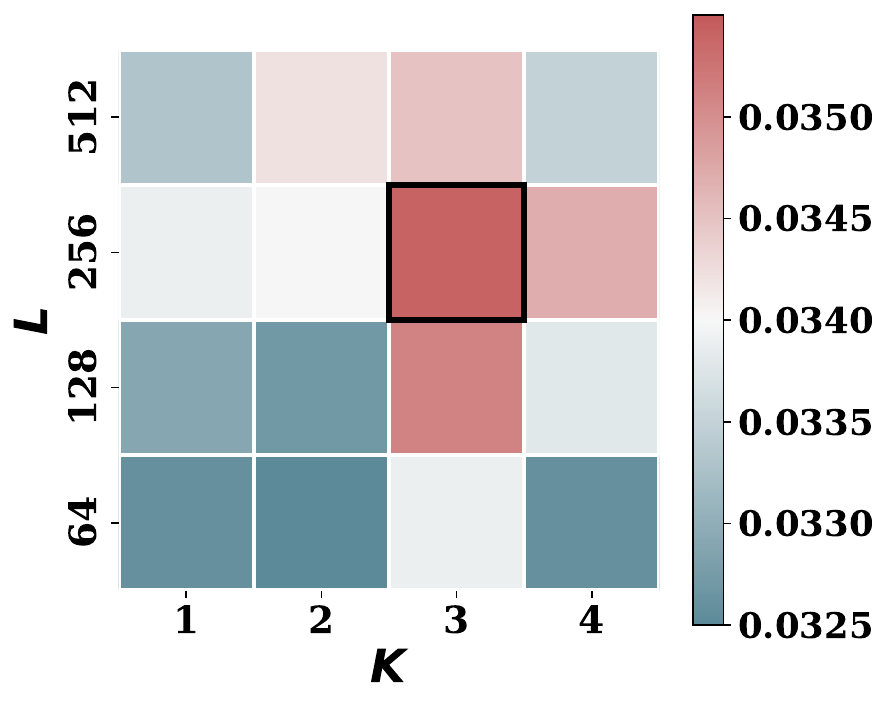}
        \caption{Amazon-Beauty-NDCG@10}
    \end{subfigure}
    \caption{ The effect of the number of sub-codebooks $K$ and the number of tokens in each sub-codebook $L$}
    \label{fig:codebook}
    \Description{codebook}
\end{figure}

We conduct a sensitivity study on the codebook configuration, examining the impact of the number of sub-codebooks $K$ and the number of tokens in each sub-codebook $L$, to examine their impact on recommendation performance. 
For fair comparison and simplicity, the top-$\mathcal{S}$ candidates for stochastic assignment are empirically set to 2.
The sensitivity results are presented in Fig.~\ref{fig:codebook}, with the optimal settings highlighted by black boxes.
The best performance is consistently achieved at $K=3$ across all three datasets, indicating that three sub-codebooks strike an effective balance between expressiveness and stability: smaller $K$ may underfit diverse collaborative semantics, while larger $K=3$ may over-fragment tokenization and undermine semantic coherence.
As for the number of codewords in each sub-codebook, the optimal $L$ varies across datasets, appearing to correlate with dataset scale and complexity. Larger datasets (e.g., Amazon-Beauty) benefit from a larger codebook size to encode richer interaction patterns, while sparser datasets (LastFM and MovieLens-1M) prefer a more moderate codebook size, where overly large codebooks as overly large codebooks may introduce redundancy and destabilize token assignment.
Finally, we note that increasing the codebook size alone results in limited performance gains, indicating that naive expansion of the discrete token space is insufficient. This highlights the importance of designing a tokenization approach that jointly models user and item tokens in a manner tailored to the diffusion-based training paradigm.

\begin{figure}[tp]
    \centering
    \begin{subfigure}[b]{0.33\textwidth}
        \includegraphics[width=\textwidth]{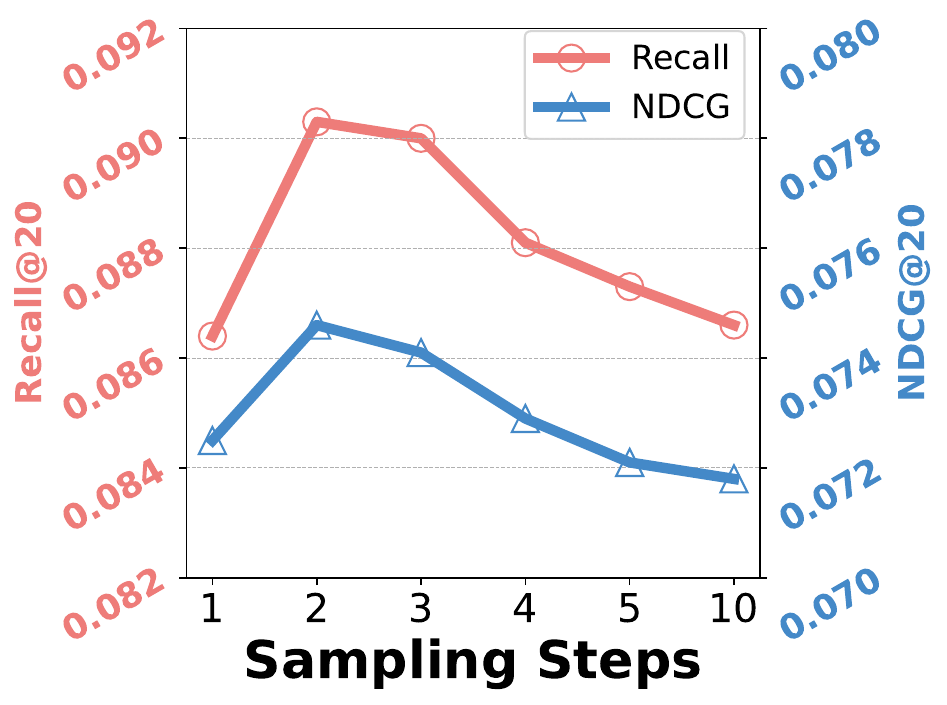}
        \caption{LastFM}
    \end{subfigure}
    \begin{subfigure}[b]{0.33\textwidth}
        \includegraphics[width=\textwidth]{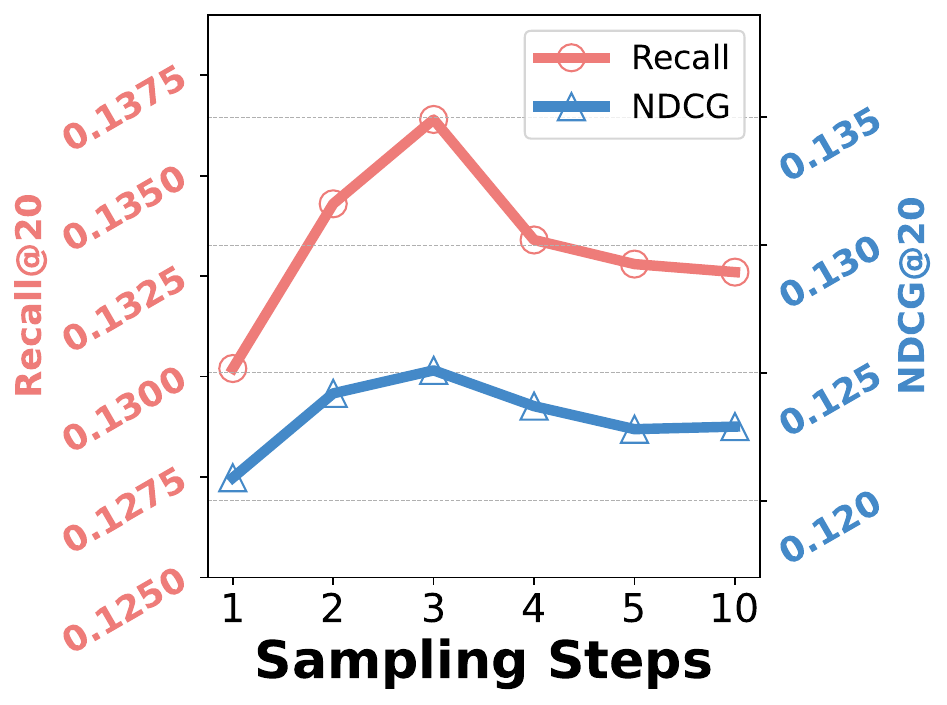}
        \caption{MovieLens-1M}
    \end{subfigure}
    \begin{subfigure}[b]{0.33\textwidth}
        \includegraphics[width=\textwidth]{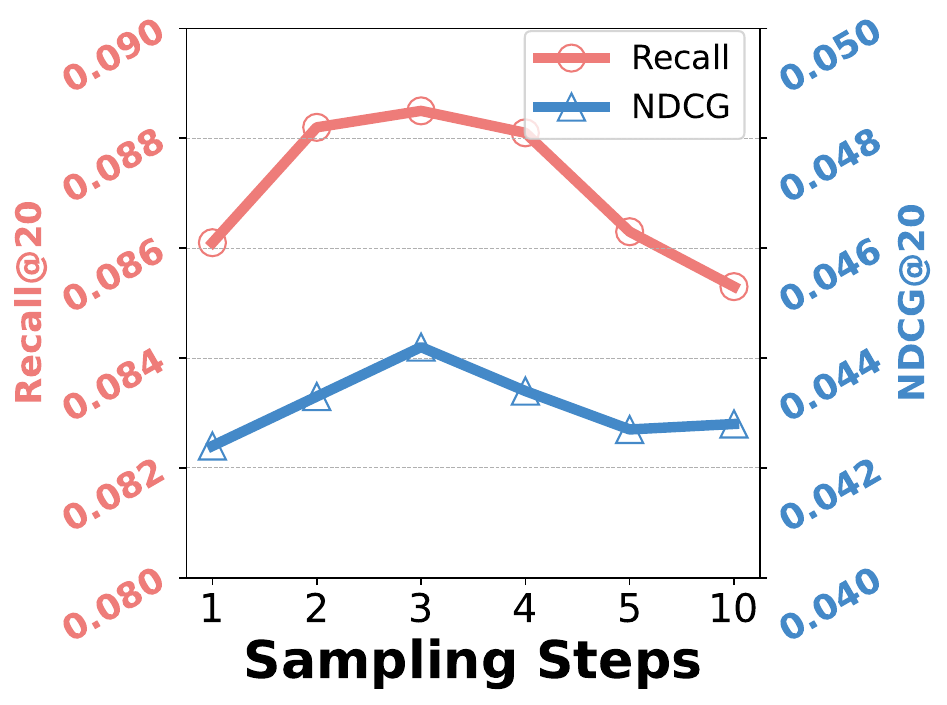}
        \caption{Amazon-Beauty}
    \end{subfigure}
    
    \begin{subfigure}[b]{0.33\textwidth}
        \includegraphics[width=\textwidth]{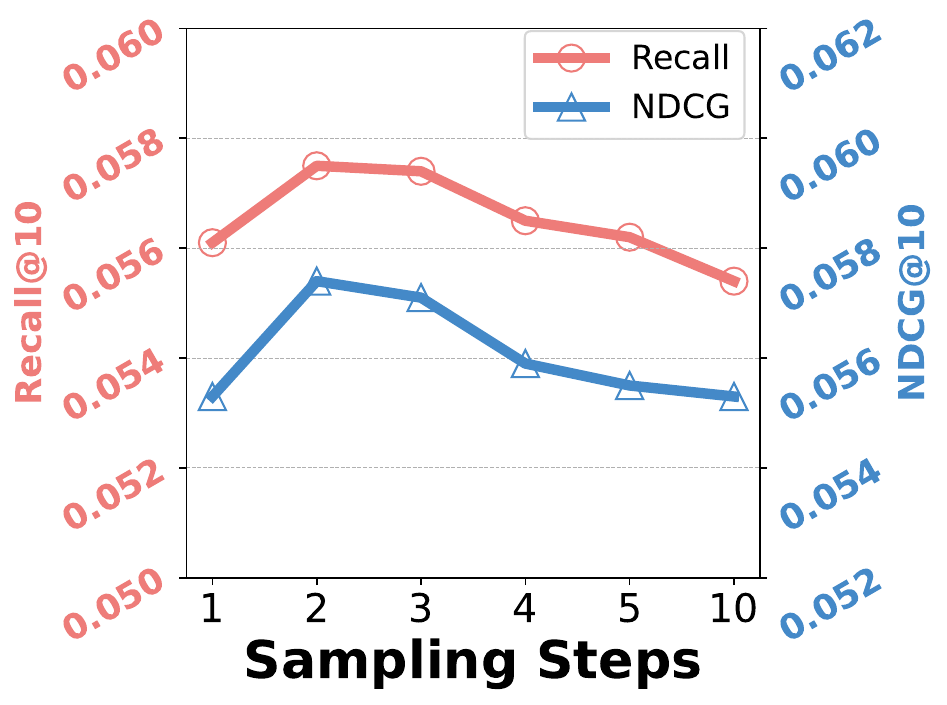}
        \caption{LastFM}
    \end{subfigure}
    \begin{subfigure}[b]{0.33\textwidth}
        \includegraphics[width=\textwidth]{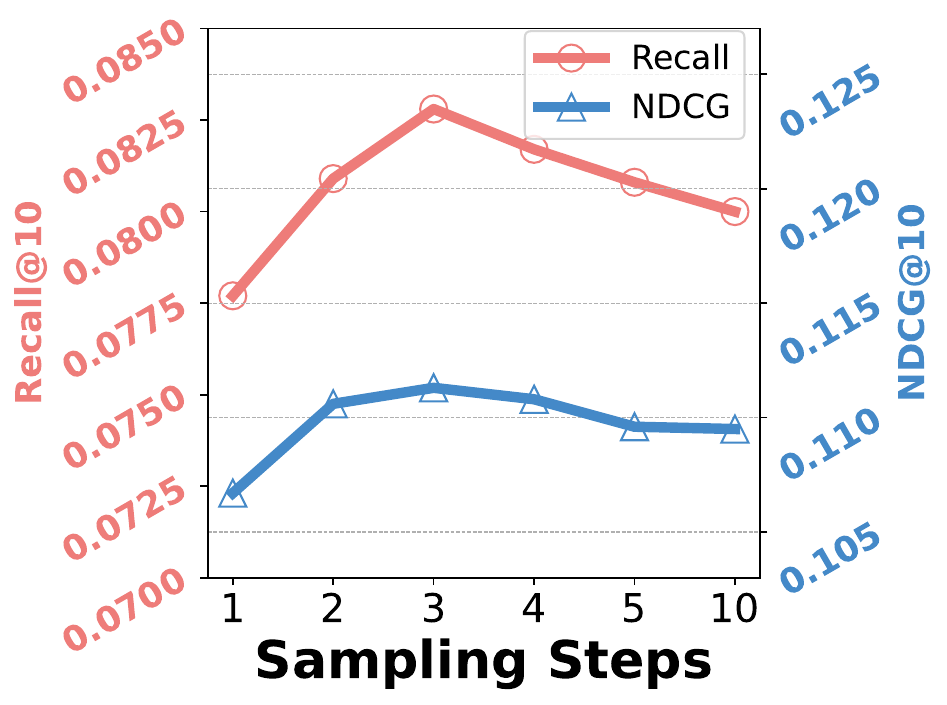}
        \caption{MovieLens-1M}
    \end{subfigure}
    \begin{subfigure}[b]{0.33\textwidth}
        \includegraphics[width=\textwidth]{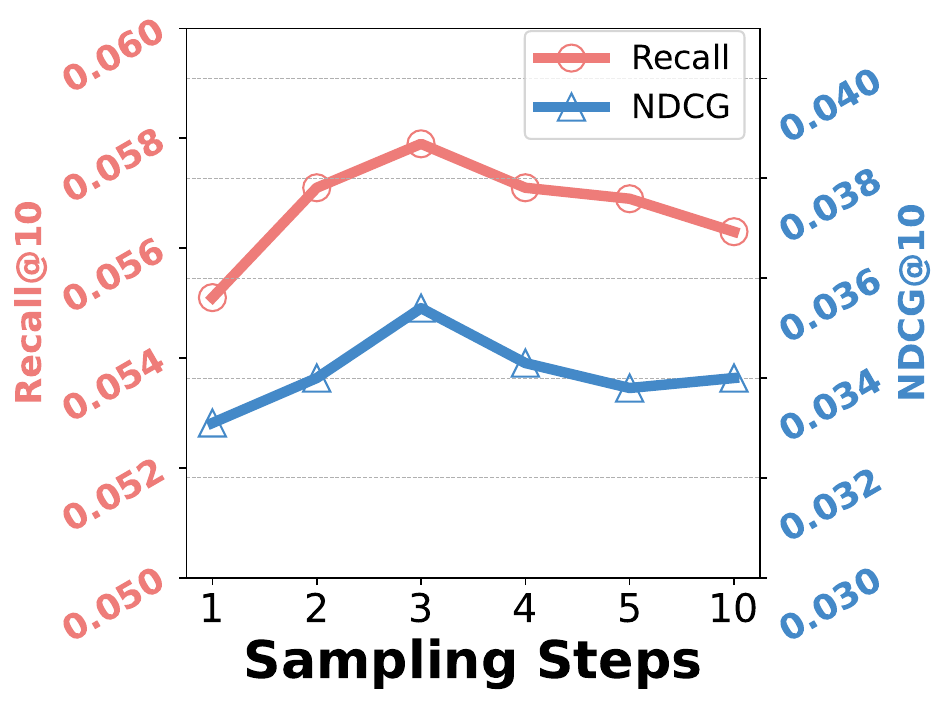}
        \caption{Amazon-Beauty}
    \end{subfigure}
    \caption{Effect of diffusion sampling steps $S$ on three datasets}
    \label{fig:sampling}
    \Description{sampling steps}
\end{figure}

\subsubsection{Effect of Diffusion Sampling Steps}
We further study the effect of the number of sampling steps during inference, with the results shown in Fig.~\ref{fig:sampling}.
Across all three datasets and for both Recall and NDCG, DLMRec reaches its optimal performance at 2 sampling steps on LastFM, and at 3 sampling steps on MovieLens-1M and Amazon-Beauty. This indicates that only a few refinement iterations are needed to effectively enhance prediction quality by leveraging bidirectional denoising and iterative correction.
However, increasing the number of sampling steps further does not continue to improve performance. Instead, excessive refinement may introduce redundant updates and perturb already stable predictions, leading to slight performance degradation. This trend is consistent with our motivation for the stability-aware refinement and voting mechanism, which is designed to preserve reliable intermediate predictions while reducing unnecessary perturbations during iterative generation.

\subsection{Model Investigation}
This section presents a comprehensive evaluation of DLMRec, including analyses of inference efficiency and qualitative case studies of CAST.

\subsubsection{Inference Time Analysis}

\begin{figure}[htp]
\centering
  \includegraphics[width=0.5\textwidth]{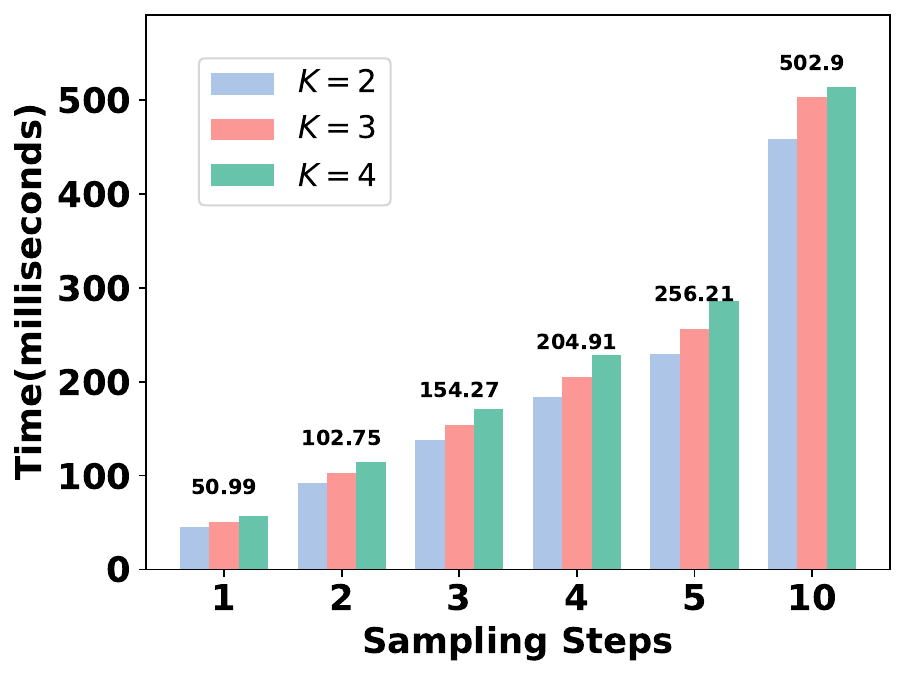}
\caption{Inference time analysis of DLMRec on MovieLens-1M under different numbers of sampling steps and item token lengths. For clarity, the exact sampling times are annotated for the case of $K=3$, which corresponds to the best-performing token setting. }
\Description{Time Analysis}
\label{fig:time} 
\end{figure}

\begin{table}[bp]
\centering
\caption{Comparison of the average inference time per user (in milliseconds) for Top-20 recommendations.}
\begin{tabular}{c|c}
\toprule
\toprule 
\textbf{Method} & \textbf{Inference Time} \\
\hline
LLaRa & 45.86 \\
CoLLM & 97.81 \\
TokenRec & 161.80 \\
DLMRec 1-step & 50.99\\
DLMRec 2-step & 102.75\\
DLMRec 3-step & 154.27\\
DLMRec 4-step & 204.91\\
\bottomrule
\bottomrule
\end{tabular}
\label{tab:sampling}
\end{table}

We analyze the inference efficiency of DLMRec on the MovieLens-1M dataset. Specifically, we measure the average time required to generate top-20 recommendations under different settings of item tokenization granularity and sampling steps. For tokenization granularity, the number of item tokens per item is varied in ${2,3,4}$, while for sampling efficiency, the number of diffusion refinement steps is set to ${1,2,3,4,5,10}$. This design allows us to systematically examine how token-level representation complexity and iterative sampling depth affect the overall generation efficiency of the proposed diffusion-based recommendation framework.
The results are shown in Fig.~\ref{fig:time}. LLaRA and CoLLM are faster than DLMRec and TokenRec, which may be attributed to their use of continuous token representations rather than discrete token generation. Compared with TokenRec under the same three-token setting, DLMRec achieves lower inference time when using a small number of sampling steps (e.g., 1--3 steps). This result suggests that diffusion-based generation can provide a more efficient alternative to autoregressive decoding under the same token setting, especially when high-quality recommendations can be obtained within a limited number of refinement steps.

We further compare the inference efficiency of DLMRec with other LLM-based recommender systems, including LLaRA, CoLLM, and TokenRec. For a fair comparison, all methods are implemented with comparable 8B-scale backbones. In addition, for TokenRec, we adopt the three-token variant to match the tokenization setting used in DLMRec.
The results are reported in Table~\ref{tab:sampling}. LLaRA and CoLLM exhibit lower inference latency than DLMRec and TokenRec, which may be attributed to their use of continuous token representations rather than discrete token generation. In contrast, under the same three-token setting, DLMRec achieves lower inference time than the autoregressive TokenRec when only a small number of sampling steps is used (e.g., 1--3 steps). This suggests that diffusion-based generation can provide a more efficient alternative to autoregressive decoding under the same token setting, especially when high-quality recommendations can be obtained within a limited number of refinement steps.

\subsubsection{CAST Case Study}

\begin{figure}[thp]
    \centering
    \begin{subfigure}[b]{0.33\textwidth}
        \includegraphics[width=\textwidth]{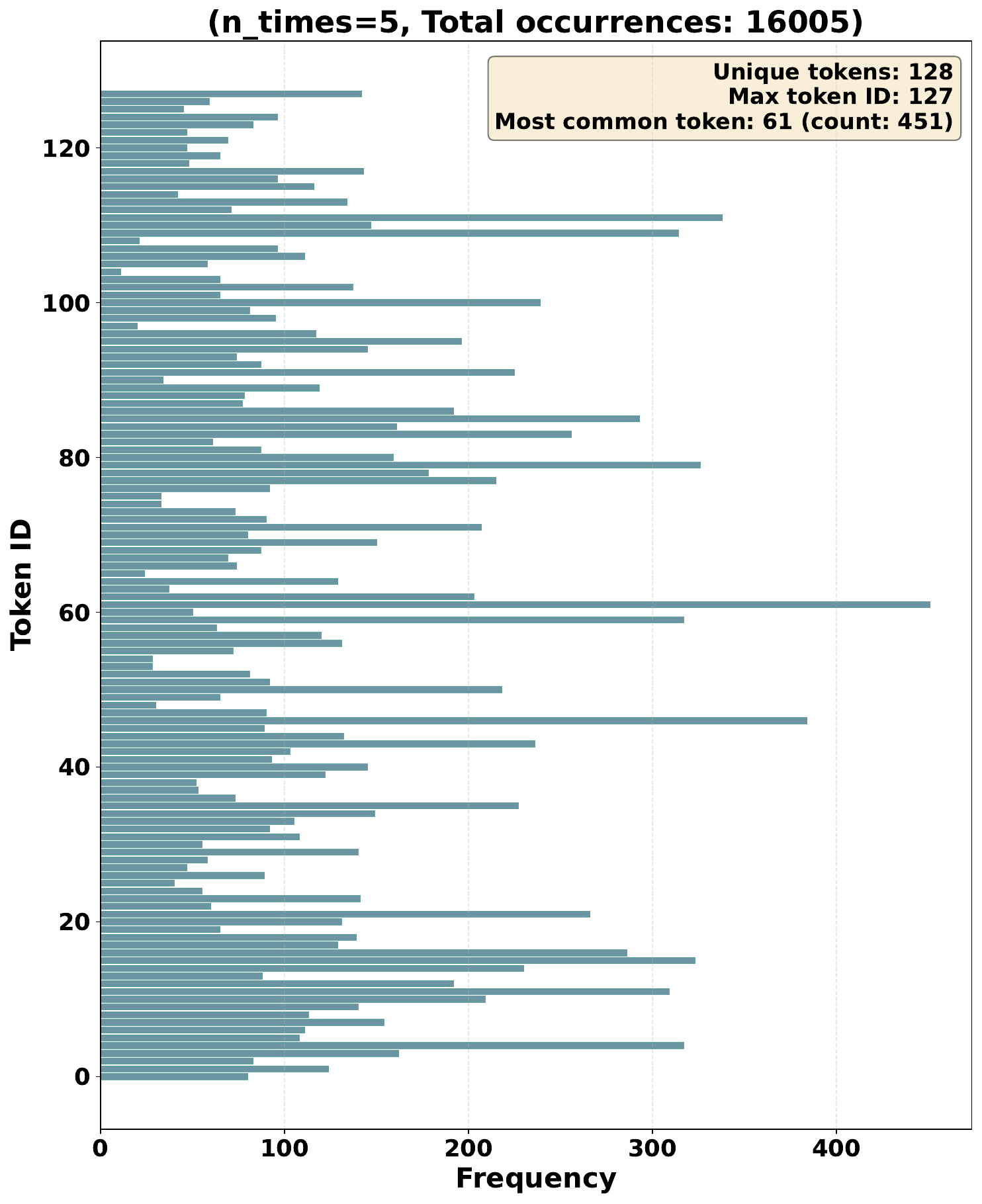}
        \caption{Token Distribution at Hop 0 for Items}
    \end{subfigure}
    \begin{subfigure}[b]{0.33\textwidth}
        \includegraphics[width=\textwidth]{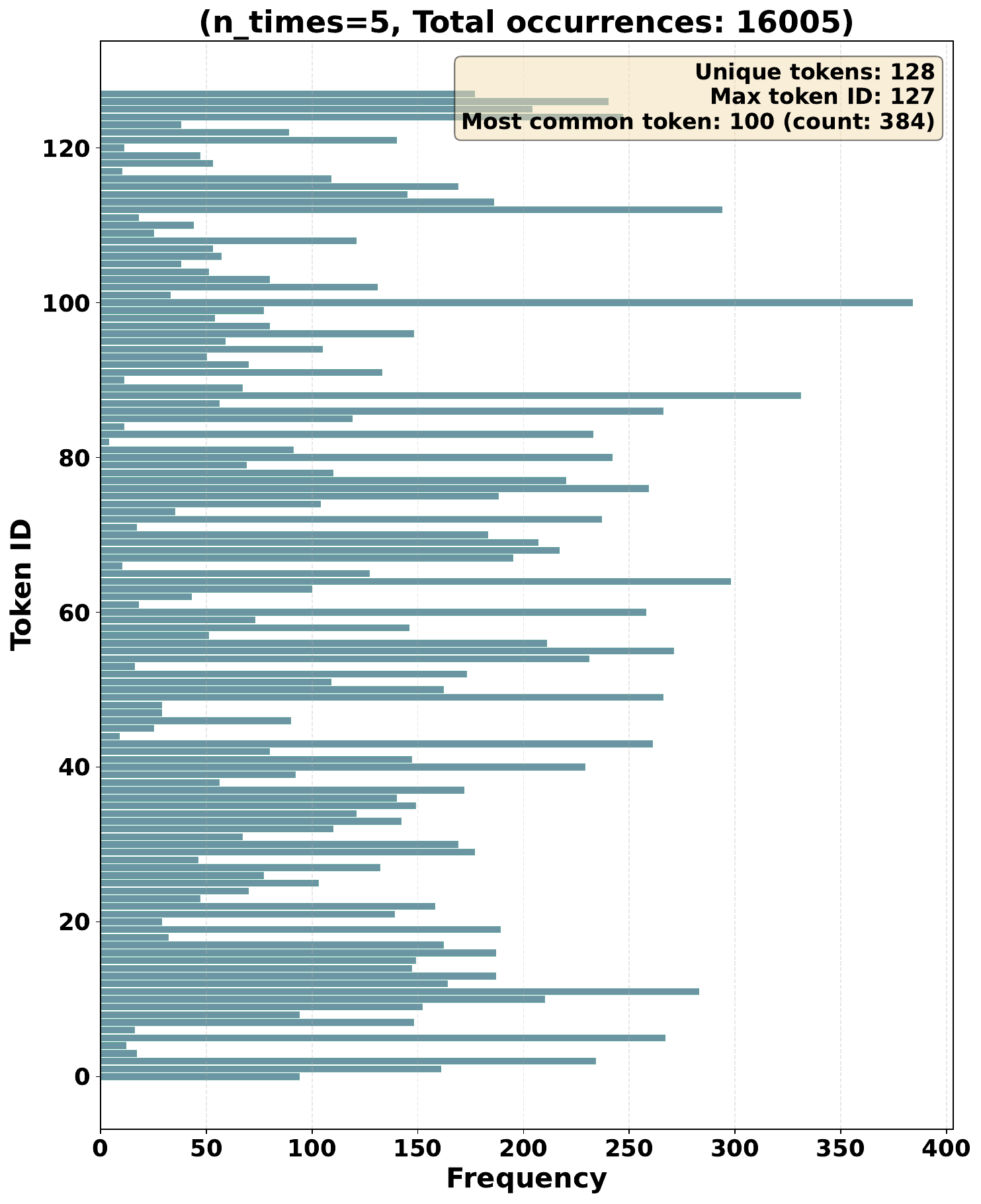}
        \caption{Token Distribution at Hop 1 for Items}
    \end{subfigure}
    \begin{subfigure}[b]{0.33\textwidth}
        \includegraphics[width=\textwidth]{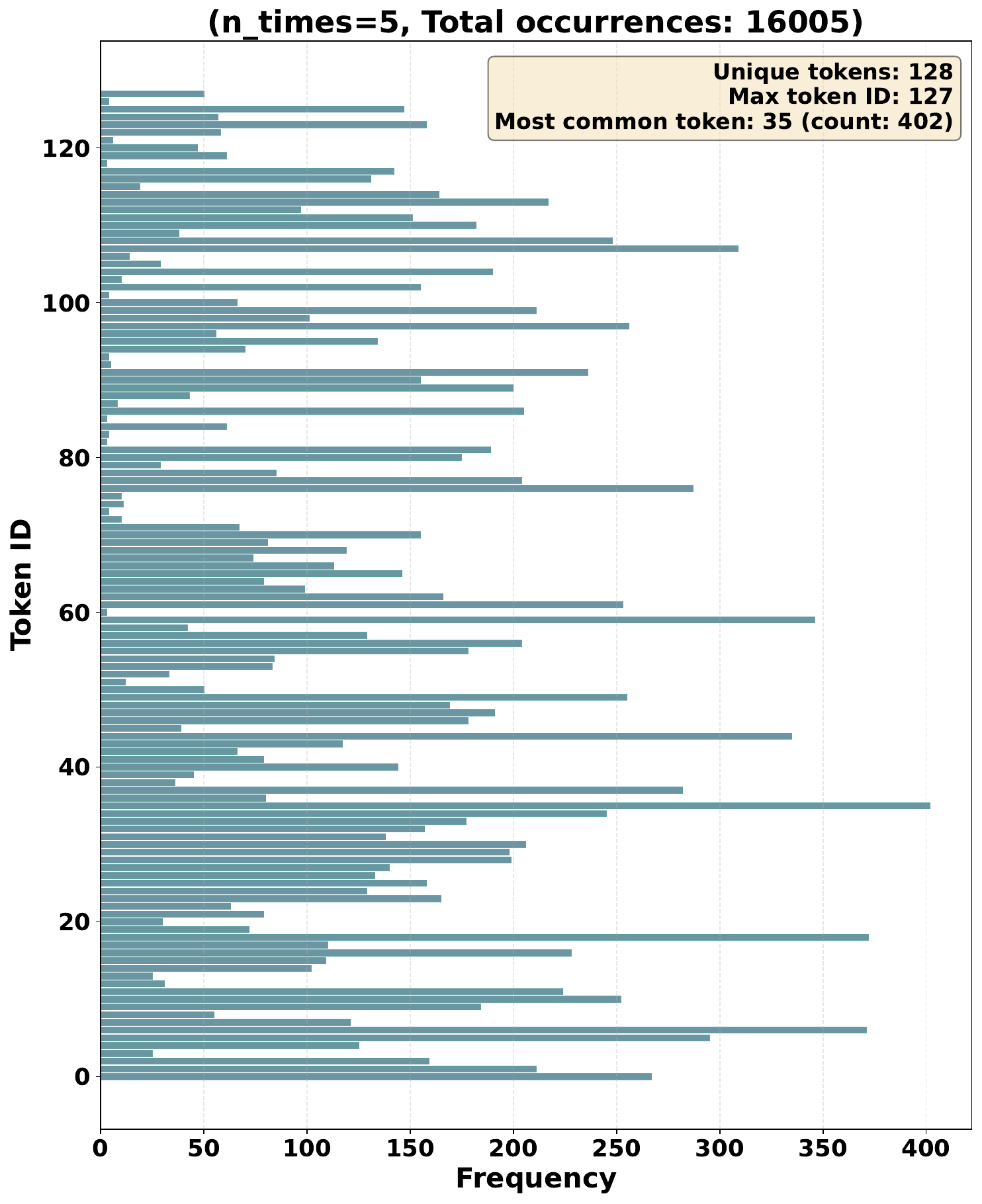}
        \caption{Token Distribution at Hop 2 for Items}
    \end{subfigure}
    \label{fig:token_dis}
    \Description{Token Distribution}
\end{figure}

To provide an intuitive understanding of the proposed tokenizer, we present a real prompt example from the MovieLens-1M dataset in Fig.~\ref{fig:prompt}.
The example corresponds to user 5159, whose interaction history includes \textit{Star Wars: Episode IV - A New Hope} and \textit{Star Wars: Episode I - The Phantom Menace}. As movies from the same series, these two items are expected to share similar collaborative semantics. Consistent with this intuition, their first-stage token is identical, i.e., \texttt{<item\_0\_72>}, while the subsequent tokens differ, enabling the model to distinguish between the two movies.
To further examine the behavior of CAST at the corpus level, we visualize the token distributions of item representations at the three hops on the MovieLens-1M dataset. For each item, stochastic token assignment is repeated $5$ times, and the resulting token frequencies are aggregated to characterize the overall usage pattern of the codebook.
Several observations can be drawn from the three distributions. First, all hops exhibit broad codebook utilization, indicating that CAST does not collapse to a few dominant codewords and preserves sufficient diversity in the discrete representation space. Second, the distributions are clearly non-uniform, suggesting that stochastic assignment is not random noise but reflects stable preferences over certain codewords induced by collaborative structure. More importantly, the three hops show distinct distributional patterns, implying that they encode different levels of collaborative semantics.
Overall, these qualitative results suggest that CAST learns diffusion-compatible discrete representations grounded in collaborative signals, preserving semantic similarity across related items while maintaining token-level discriminability.

%% file: tables/dataset_stats.tex
\begin{table}[htp]
\centering
\caption{Statistics of the datasets}
\begin{tabular}{c|ccc}
\toprule
\toprule 
\textbf{Dataset} & \textbf{LastFM} & \textbf{MovieLens-1M} & \textbf{Amazon-Beauty}\\
\hline
\#Users & 1,865 & 6,035 & 18,028\\
\#Items & 3,517 & 3,201 & 10,266 \\
\#Interactions & 74,154 & 574,684 & 138,527 \\
Interaction Density & 1.1305\%  & 2.9748\% & 0.0748\% \\
\bottomrule
\bottomrule

\end{tabular}
\label{tab:dataset_stats}
\end{table}

%% file: tables/perf.tex
\begin{table*}[tp]
\caption{Summarization of the overall performance comparison across three recommendation datasets. The best performance is highlighted in bold, and the second best is underlined. \%Improve denotes the relative improvement of DLMRec over the strongest baseline.}
\centering
\begin{adjustbox}{width=\textwidth}
\begin{tabular}{ccccccccccccc}
\toprule
\toprule
\textbf{Dataset}  & \multicolumn{4}{c}{\textbf{LastFM}} & \multicolumn{4}{c}{\textbf{MovieLens-1M}} & \multicolumn{4}{c}{\textbf{Amazon-Beauty}} \\
\textbf{Method}   & \textbf{R@20} & \textbf{R@10} & \textbf{N@20} & \textbf{N@10} & \textbf{R@20} & \textbf{R@10} & \textbf{N@20} & \textbf{N@10} & \textbf{R@20} & \textbf{R@10} & \textbf{N@20} & \textbf{N@10} \\
\hline
\multicolumn{13}{c}{\cellcolor{gray!20}\textit{\textbf{GNN-based Collaborative Filtering}}} \\
LightGCN & 0.0595 & 0.0311 & 0.0425 & 0.0297 & 0.1088 & 0.0620 & 0.1062 & 0.0984 & 0.0730 & 0.0459 & 0.0349 & 0.0268\\
SGL & 0.0612 & 0.0347 & 0.0431 & 0.0329 & 0.1133 & 0.0627 & 0.1113 & 0.0946 & 0.0741 & 0.0465 & 0.0353 & 0.0272\\
LTGNN  & 0.0631 & 0.0363 & 0.0452 & 0.0345 & 0.1173 & 0.0633 & 0.1144 & 0.0992 & 0.0768 & 0.0474 & 0.0375 & 0.0283\\
\multicolumn{13}{c}{\cellcolor{gray!20}\textit{\textbf{Sequential Recommendation }}} \\
SASRec & 0.0587 & 0.0308 & 0.0425 & 0.0292 & 0.1021 & 0.0587 & 0.1001 & 0.882 & 0.0724 & 0.0457 & 0.0457 & 0.0328\\
BERT4Rec & 0.0571 & 0.0303 & 0.0416 & 0.0285 & 0.1023 & 0.0598 & 0.1007 & 0.0899 & 0.0736 & 0.0499 & 0.0383 & 0.0311\\
\multicolumn{13}{c}{\cellcolor{gray!20}\textit{\textbf{Diffusion-based Generative Recommendation }}} \\
DiffRec & 0.0606 & 0.0293 & 0.0429 & 0.0273 & 0.1171 & 0.0641 & 0.1080 & 0.0997 & 0.0702 & 0.0497 & 0.0373 & 0.0310\\
CDRec & 0.0693 & 0.0431 & 0.0561 & 0.0420 & \underline{0.1276} & 0.0742 & 0.1168 & 0.1014 & 0.0767 & 0.0506 & 0.0376 & 0.0298\\
LLaDaRec  & 0.0747 & 0.0524 & 0.0704 & \underline{0.0542} & 0.1196 & 0.0771 & 0.1179 & 0.1033 & 0.0754 & 0.0523 & 0.0409 & \underline{0.0339}\\
\multicolumn{13}{c}{\cellcolor{gray!20}\textit{\textbf{LLM-
empowered Recommenders }}} \\
TIGER & 0.0718 & 0.470 & 0.0639 & 0.0504 & 0.1052 & 0.0751 & 0.1096 & 0.1025 & 0.0746 & 0.0519 & 0.0381 & 0.0318\\
LLaRa & \underline{0.0858} & \underline{0.0535} & 0.0708 & 0.0538 & 0.1252 & \underline{0.0783} & \underline{0.1187} & \underline{0.1078} & \underline{0.0828} & \underline{0.0535} & 0.0383 & 0.0306\\
CoLLM & 0.0845 & 0.0458 & \underline{0.0713} & 0.0499 & 0.1229 & 0.0735 & 0.1061 & 0.1012 & 0.0815 & 0.0532 & \underline{0.0411} & 0.0326\\
TokenRec & 0.0836 & 0.0469 & 0.0682 & 0.0479 & 0.1216 & 0.0685 & 0.1144 & 0.1021 & 0.0811 & 0.0517 & 0.0375 & 0.0294\\
\rowcolor{red!13} 
DLMRec &  \textbf{0.0903} & \textbf{0.0575} & \textbf{0.0746} & \textbf{0.0574} & \textbf{0.1364} & \textbf{0.0828} & \textbf{0.1251} & \textbf{0.1113} & \textbf{0.0885} & \textbf{0.0579}& \textbf{0.0442} & \textbf{0.0354} \\
\rowcolor{blue!13}
\%Improve & 5.24\% & 7.48\% & 4.63\% & 5.90\% & 6.90\% & 5.75\% & 5.39\% & 3.25\% & 6.88\% & 8.22\% & 7.54\% & 4.42\% \\
\bottomrule
\bottomrule
\end{tabular}
\end{adjustbox}

\label{OverallPerfromance}
\end{table*}

%% file: sections/Related_work.tex
\section{Related Work}
\label{Related_work}

This section reviews three lines of related work closely connected to our study: diffusion language models, diffusion-based recommender models, and LLM-based recommender systems.

\subsection{Diffusion Language Models}

Recently, diffusion language models have emerged as a promising alternative to autoregressive language modeling. Instead of generating tokens strictly in a left-to-right manner, they model language through iterative denoising, demonstrating advantages in bidirectional context modeling, global coherence, and flexible generation ~\cite{li2023diffusion, li2025survey, he2026graph}.

\subsubsection{Continuous-state Diffusion Language Models} 

Driven by the success of continuous diffusion models in image synthesis, early diffusion language models typically projected discrete tokens into continuous embeddings and performed diffusion in latent continuous space.
Among the pioneering works in this line, Diffusion-LM ~\cite{li2022diffusion} aims to generate word embeddings from Gaussian noise and converts them back to discrete tokens through a rounding function.
Building on this embedding-based paradigm, DiffuSeq ~\cite{gong2022diffuseq} further extends the continuous diffusion to conditional sequence generation under a classifier-free latent-space framework. 
In a different direction, SSD-LM~\cite{han2023ssd} performs diffusion over vocabulary-level simplex representations rather than learned latent embeddings, enabling semi-autoregressive, block-wise generation. More recently, RDLM~\cite{jo2026continuous} revisits continuous-state diffusion language modeling from a geometric perspective. By modeling continuous flow on the statistical manifold, it substantially improves continuous diffusion modeling for language and further narrows the gap to autoregressive models.

\subsubsection{Discrete-state Diffusion Language Models}

There has been growing interest in discrete diffusion models for their capacity to learn complex data distributions in discrete spaces~\cite{xu2024discrete}.
D3PM~\cite{austin2021structured} introduced the first discrete-state diffusion framework for categorical variables. Replacing Gaussian noise with discrete transition operators, it defines several stationary corruption processes, including uniform diffusion, where transitions to all states are equally likely, and absorbing state diffusion, where tokens are progressively replaced by a special mask token.
Based on the absorbing state, DiffusionBERT ~\cite{he2023diffusionbert} intoduces a spindle noise schedule to leverage the domain knwoledge, enhancing both generation quality and training efficiency. 
\citet{campbell2022continuous} were the first to cast discrete-state diffusion in continuous time through stochastic differential equations (SDEs), thereby improving the flexibility of the sampling process. However, the reverse process remains discretely parameterized, which limits its empirical performance. \citet{sun2022score} subsequently proposed ratio matching to learn marginal probabilities via maximum-likelihood training. Building on this line, later methods such as SEDD~\cite{lou2024discrete} developed alternative score-based formulations for discrete diffusion modeling.

More recently, diffusion language models have begun to scale toward the large language model regime. LLaDA~\cite{nie2025large} extends masked diffusion language modeling to 8B parameters under the standard pre-training and supervised fine-tuning pipeline, demonstrating competitive performance in scalability, in-context learning, and instruction following. Building upon this direction, Seed Diffusion~\cite{song2025seed} further highlights the efficiency advantage of diffusion-based generation through block-level parallel sampling. Subsequently, Dream~\cite{ye2025dream} improves diffusion-based LLMs via autoregressive-model initialization and context-adaptive token-level noise rescheduling. Moreover, Mercury~\cite{khanna2025mercury} focuses on commercial-scale diffusion LLMs with ultra-fast inference, demonstrating their potential for latency-sensitive real-world applications.
These advances position diffusion language models as a competitive large-scale paradigm. Their bidirectional and iterative refinement properties present a promising avenue for recommendation research.

\subsection{Diffusion Recommender Models}

Diffusion models, as a powerful generative paradigm, have recently achieved remarkable success in recommendation, substantially advancing a variety of recommendation tasks through their effective generation capabilities ~\cite{zhu2024graph, wu2026beyond}.

\subsubsection{Continuous-state Diffusion Recommender Models}

DiffRec~\cite{wang2023diffusion} represents one of the earliest attempts to incorporate diffusion modeling into recommendation systems, where Denoising Diffusion Probabilistic Model~\cite{ho2020denoising} is applied to binary interaction vectors for user-item preference modeling.
Building upon this foundation, GiffCF~\cite{zhu2024graph} introduces graph-based diffusion mechanisms to model higher-order collaborative relations, while BSPM~\cite{choi2023blurring} formulates recommendation over the adjacency matrix within a continuous-time diffusion framework.
From a different perspective, DreamRec~\cite{yang2023generate} casts sequential recommendation in a learn-to-generate paradigm by applying diffusion to target item representations conditioned on users’ historical interactions.
Furthermore, PreferDiff~\cite{liu2024preference} incorporates negative samples into diffusion-based sequential recommendation, establishing a closer connection between diffusion modeling and preference-oriented learning objectives.

\subsubsection{Discrete-state Diffusion Recommender Models}

DCDR~\cite{lin2024discrete} pioneers the use of discrete-state diffusion in recommendation for reranking tasks. It perturbs item positions within interaction sequences, enabling diffusion-based learning directly over discrete collaborative signals.
Beyond reranking, DDSR~\cite{xie2024breaking} extends discrete-state diffusion to sequential recommendation by introducing semantic IDs to capture the uncertainty and fuzziness of user interests.
More recently, PreferGrow~\cite{hu2026fading} advances this line by modeling preference ratios between item pairs, where user preferences are faded through item replacement and reconstructed by iteratively growing preference signals from the estimated ratios. CDRec~\cite{liu2026continuous} further explores a continuous-time discrete-space diffusion framework for recommendation, leveraging absorbing-state diffusion and domain-aware noise scheduling to improve both recommendation accuracy and sampling efficiency.
LLaDA-Rec~\cite{shi2025llada} applies discrete diffusion to parallel semantic ID generation, introducing a parallel tokenization scheme and two masking strategies to capture both inter-item sequential dependencies and intra-item semantic relationships.
Despite these advances, most existing methods focus on recommendation-specific diffusion modeling, while the potential of large language models for richer semantic understanding and generation remains underexplored.

\subsection{LLM-based Recommender Systems}

Recently, large language models (LLMs) have been widely adopted to enhance recommender systems due to their strong capabilities in semantic understanding, reasoning, and generative modeling. However, effectively applying LLMs to recommendation remains challenging because of the intrinsic gap between natural language modeling and recommendation-oriented preference modeling~\cite{cikm25-gen-rec-tutorial, ning2026efficiency}.

\subsubsection{Text-based Alignment for LLM-based Recommendation}

Early LLM-based recommendation methods mainly rely on text-based alignment, which reformulates user-item interactions into natural language sequences through prompts and instruction tuning. 
For example, TALLRec~\cite{bao2023tallrec} converts user histories and item metadata into instruction-following data, enabling LLMs to infer user preferences from textual contexts. 
Moreover, P5~\cite{geng2022recommendation} extends this paradigm by unifying multiple recommendation tasks within a text-to-text framework, where interactions, metadata, and recommendation signals are transformed into natural language sequences processed through personalized prompts and whole-word embeddings for tokenized user and item IDs.

\subsubsection{Representation-based Alignment with Collaborative Signals}

Representation-based methods leverage continuous encodings of collaborative information to facilitate the integration of language modeling and preference learning.
For instance, LLaRA~\cite{liao2024llara} combines textual item metadata with item representations learned from traditional sequential recommenders, and projects these embeddings into the LLM input space through a trainable projector. Building upon this idea, CoLLM~\cite{zhang2025collm} further incorporates user and item collaborative representations as an additional modality, enabling LLM-based recommendation to jointly exploit textual semantics and collaborative signals.
Moving beyond directly injecting user and item representations, ContRec~\cite{qu2025diffusion} constructs continuous tokens from collaborative information and introduces a diffusion module for implicit user preference modeling. 
From a different perspective, E4SRec~\cite{li2023e4srec} revisits LLM-based recommendation under an ID-only paradigm. Instead of relying on textual semantics, it injects pretrained item ID embeddings into LLMs to enable efficient and controllable sequential recommendation.

\subsubsection{Token-based Alignment for LLM-based Recommendation}

Token-based methods represent users and items as discrete tokens to enable LLM-based recommendations. 
CLLM4Rec~\cite{zhu2024collaborative} directly extends the LLM vocabulary with user and item ID tokens, learning their collaborative semantics through soft and hard prompting. 
LC-Rec~\cite{zheng2024adapting} further learns semantic item indices via vector quantization and alignment tuning, enabling direct generation from the full item vocabulary. 
Moreover, TokenRec~\cite{qu2025tokenrec} incorporates high-order collaborative knowledge through masked vector quantization, strengthening the integration of graph-based signals with discrete representations.

%% file: sections/Conclusion.tex
\section{Conclusion}
\label{Conclusion}

In this paper, we propose DLMRec, a diffusion language modeling framework tailored for recommendation. Departing from the dominant autoregressive paradigm, DLMRec reformulates recommendation as a discrete denoising process, enabling bidirectional context modeling and iterative whole-sequence refinement for next-item prediction. To adapt diffusion language modeling to recommendation, we develop a collaborative-aware stochastic tokenizer that converts multi-hop collaborative signals into diffusion-compatible discrete tokens, a curriculum-driven training strategy that progressively aligns denoising with preference learning at both the item and token levels, and a stability-aware voting mechanism that aggregates evidence across refinement steps for more consistent final prediction. Extensive experiments on three real-world datasets demonstrate that DLMRec consistently outperforms strong baselines, validating the effectiveness of diffusion-based recommendation modeling. These results suggest that diffusion language models provide a promising new paradigm for recommender systems beyond autoregressive generation.